\documentclass[11pt]{article}

\usepackage{amsmath}  

\usepackage{hyperref}
\usepackage{url}

\usepackage{amssymb}  
\usepackage{float} 
\usepackage{xcolor} 
\usepackage{graphicx} 
\usepackage{cite} 
\usepackage[normalem]{ulem}
\usepackage{bbm}
\usepackage{dsfont}
\usepackage[mathscr]{euscript}
\usepackage{eufrak}
\usepackage{ntheorem}

\usepackage{cite}

\usepackage{cite}

\usepackage{varwidth}

\usepackage[normalem]{ulem}
\usepackage{tensor}
\usepackage{emptypage}
\usepackage{slashed}
\usepackage{color}
\usepackage{graphicx}  
\usepackage{psfrag}
\usepackage{tocbibind} 
\usepackage{epsf}
\usepackage{epic}
\usepackage{eepic}
\usepackage{epsfig}
\usepackage{wrapfig}
\usepackage{a4}
\usepackage{amssymb,latexsym}
\usepackage[active]{srcltx}
\usepackage{url}
\usepackage{fancybox}
\usepackage{ntheorem}
\usepackage{mhequ}
\usepackage[g]{esvect}
\usepackage{epstopdf}

\usepackage{wasysym}

\usepackage{multirow}

\usepackage{tikz}  \usetikzlibrary{patterns}

\usepackage{bbm}
\usepackage{slantsc}
\usepackage{slashed}

\usepackage[normalem]{ulem}

\usepackage{marginnote}

\newcommand{\mcM}{{\mycal M}}
\newcommand{\mcD}{{\mycal D}}

\newcommand{\bmcM}{ {\color{red} \,\,\,\,\widetilde{\!\!\!\!\mcM}}}

\newcommand{\omcM}{\bmcM}
\newcommand{\ovmcM}{\omcM}

\newcommand{\bchi}{\blue{\mathbf{\chi}}}

\newcommand{\wtthyp}{\red{\,\,\,\widetilde{\!\!\!\hyp}}\!}
\newcommand{\wthyp}{\red{\,\,\,\widetilde{\!\!\!\hyp}}}
\newcommand{\whhyp}{\red{\,\,\,\widehat{\!\!\!\hyp}}\!}
\newcommand{\hthyp}{\whhyp}
\newcommand{\whyp}{\whhyp}

\newcommand{\slope}{\blue{\delta}}
\newcommand{\gap}{\blue{\hat \delta}}

\renewcommand{\hbar}{{\overline h}}%

\newcommand{\backriemg}{\red{\mathfrak{g}}}

\newcommand{\divM}{\blue{\mathrm{div}_\hyp}}
\newcommand{\gradM}{\blue{\nabla^\hyp}}

\newcommand{\ovhyp}{\ohyp}
\newcommand{\ovhypone}{\,\,\overline{\!\!\hyp_{H_1}}}

\newcommand{\hypone}{\hyp_{H_1}}

\newcommand{\ohyp}{\,\,\overline{\!\!\hyp}} 
\newcommand{\thyp}{\,\,\widetilde{\!\!\hyp}\,}

\newcommand{\oell}[1]{\overset{\mbox{\tiny ($\ell$)}}{#1}}
\newcommand{\oellp}[1]{\overset{\mbox{\tiny ($\ell+1$)}}{#1}}
\newcommand{\onp}[1]{\overset{\mbox{\tiny ($n+1$)}}{#1}}
\newcommand{\on}[1]{\overset{\mbox{\tiny ($n$)}}{#1}}
\newcommand{\onlog}[1]{\overset{\mbox{\tiny ($n+1,\log$)}}{#1}}

\newcommand{\oellpp}[1]{\overset{\mbox{\tiny ($\ell+2$)}}{#1}}

\newcommand{\oone}[1]{\overset{\mbox{\tiny (1)}}{#1}}

\newcommand{\ozero}[1]{\overset{\mbox{\tiny (0)}}{#1}}

\newcommand{\tScri}{\blue{\mathcal{I}}}

\newcommand{\blue}[1]{{\color{blue}#1}}
\newcommand{\red}[1]{{{\color{red}#1}}}

\newcommand{\secEdbh}[1]{\mnote{ptcSecEdbh environement, for a second edition in the black hole book, but this means that the stuff around has been used in the black hole notes, so the whole file should be discarded or replaced by a reference?}}

\newcommand{\ptcSecEd}[1]{\mnote{ptcSecEd environement, but this means that this has been used in the Vienna lecture notes, so the whole file should be discarded or replaced by a reference?}}

\newcommand{\tooslow}[1]{\ptc{tooslow environement; figures commented out because too long to display and compile, trying to reduce them with acrobat did not give anything useful; should be fixed in the cleanrun file}}

\newcommand{\rnoblackholesc}[2]{}
\newcommand{\waehr}[1]{}
\newcommand{\licht}[1]{}

\newcommand{\docendnp}{\ptcr{remove end document}\newpage\input{end}\end{document}}
\newcommand{\docend}{\ptcr{remove end document}\newpage\input{endWithProblems}\end{document}}

\newcommand{\ptcn}[1]{\ptcr{#1}}

\newcommand{\syncx}[1]{\ptcxx{here starts a syncx, or possibly a restrict environment}}
\renewcommand{\syncx}[1]{\ptc{syncx command here}{\color{red}#1}}

\newcommand{\roscoff}[1]{\ptc{roscoff here}}

\newcommand{\ptcxx}[1]{\mnote{{\bf ptcss:} {\color{red} #1}}}

\newcommand{\mnotex}[1]
{\protect{\stepcounter{mnotecount}}$^{\mbox{\footnotesize
$
\bullet$\themnotecount}}$ \marginpar{
\raggedright\tiny\em
$\!\!\!\!\!\!\,\bullet$\themnotecount: #1} }

\newcommand{\jamesx}[1]{}
\renewcommand{\jamesx}[1]{{\mnote{{\color{blue}{\bf jg:}
#1} }}}

\newcommand{\fourg}{\blue{\mathbf{g}}}

\newcommand{\ptcheck}[1]{\ptc{checked on #1}}

\def\ben{\begin{equation}}
\def\een{\end{equation}}
\def\bena{\begin{eqnarray}}
\def\eena{\end{eqnarray}}

\def\f(#1/#2){\frac{#1}{#2}}
\def\Frac(#1/#2){\left(\frac{#1}{#2}\right)}
\def\chris(#1-#2-#3){{\mit \Gamma}^{#1}{}_{{#2}{#3}} }
\def\tilchris(#1-#2-#3){\tilde{{\mit \Gamma}}^{#1}{}_{{#2}{#3}}}
\def\hatchris(#1-#2-#3){\hat{{\mit \Gamma}}^{#1}{}_{{#2}{#3}}}

\newcommand{\Ric}{\blue{\mathrm{Ric}\,}}

{\catcode `\@=11 \global\let\AddToReset=\@addtoreset}
\AddToReset{figure}{section}

\DeclareFontFamily{OT1}{rsfs}{}
\DeclareFontShape{OT1}{rsfs}{m}{n}{ <-7> rsfs5 <7-10> rsfs7 <10-> rsfs10}{}
\DeclareMathAlphabet{\mycal}{OT1}{rsfs}{m}{n}

{\catcode `\@=11 \global\let\AddToReset=\@addtoreset}
\AddToReset{equation}{section}

\newcounter{mnotecount}[section]

\renewcommand{\themnotecount}{\thesection.\arabic{mnotecount}}

\newcommand{\mnote}[1]
{\protect{\stepcounter{mnotecount}}$^{\mbox{\footnotesize
$
\bullet$\themnotecount}}$ \marginpar{
\raggedright\tiny\em
$\!\!\!\!\!\!\,\bullet$\themnotecount: #1} }

\newcommand{\ptcr}[1]{{\color{red}\mnote{{\color{red}{\bf ptc:}#1} }}}
\newcommand{\jj}[1]%
{{\color{red}\mnote{{\color{red}{\bf jj:} #1} }}}

\definecolor{HP}{rgb}{1,0.09,0.58}

\newcommand{\hs}{\cH_{\mbox{\scriptsize sing}}}

\newcommand{\beadl}[1]{\begin{deqarr}\label{#1}}
\newcommand{\eeadl}[1]{\arrlabel{#1}\end{deqarr}}%

\def\nz{\ifmmode {I\hskip -3pt N} \else {\hbox {$I\hskip -3pt N$}}\fi}
\def\zz{\ifmmode {Z\hskip -4.8pt Z} \else
       {\hbox {$Z\hskip -4.8pt Z$}}\fi}
\def\qz{\ifmmode {Q\hskip -5.0pt\vrule height6.0pt depth 0pt
       \hskip 6pt} \else {\hbox
       {$Q\hskip -5.0pt\vrule height6.0pt depth 0pt\hskip 6pt$}}\fi}
\def\rz{\ifmmode {I\hskip -3pt R} \else {\hbox {$I\hskip -3pt R$}}\fi}
\def\cz{\ifmmode {C\hskip -4.8pt\vrule height5.8pt\hskip 6.3pt} \else
       {\hbox {$C\hskip -4.8pt\vrule height5.8pt\hskip 6.3pt$}}\fi}
\def\au{{\setbox0=\hbox{\lower1.36775ex\hbox{''}\kern-.05em}\dp0=.36775ex\hs
kip0pt\box0}}
\def\ao{{}\kern-.10em\hbox{``}}

\newcommand\Gregbeq{\begin{eqnarray}}
\newcommand\Gregeeq{\end{eqnarray}}

\newcommand{\scri}{{\mycal I}}%

\def\cH{{\cal H}}

\def\h1{{\hat 1}}
\def\h2{{\hat 2}}

\def\3f{\frac{3}{2}}

\newcommand{\oversetty}[2]{%
\mathop{#2}\limits^{\vbox to -.1ex{%
\kern -1.5ex\hbox{$\scriptstyle #1$}\vss}}}

\newcommand{\jlcax}[1]{}
\newcommand{\eean}{\nonumber\end{eqnarray}}

\newcommand{\kk}[1]{}

\newcommand{\beq}{\begin{equation}}

\newcommand{\rgc}[1]{}

\newcommand{\FS}       
                  {F}

\newcommand{\HS} 
       {H_{\mbox{\scriptsize volume}}}

{\ptc{this should be removed in the oberwolfach version}}%

\newcommand{\eel}[1]{\label{#1}\end{equation}}
\newcommand{\eeal}[1]{\label{#1}\end{eqnarray}}
\newcommand{\bed}{\begin{deqarr}}
\newcommand{\eed}{\end{deqarr}}
\newcommand{\bedl}[1]{\begin{deqarr}\label{#1}}
\newcommand{\eedl}[2]{\arrlabel{#1}\label{#2}\end{deqarr}}

\newcommand{\mcU}{{\mycal U}}

\newcommand{\bel}[1]{\begin{equation}\label{#1}}
\newcommand{\bea}{\begin{eqnarray}}
\newcommand{\bean}{\begin{eqnarray}\nonumber}
\newcommand{\beal}[1]{\begin{eqnarray}\label{#1}}
\newcommand{\eea}{\end{eqnarray}}

\newcommand{\qed}{\hfill $\Box$}
\newcommand{\qedskip}{\hfill $\Box$\medskip}
\newcommand{\proof}{\noindent {\sc Proof:\ }}
\newcommand{\be}{\begin{equation}}
\newcommand{\eeq}{\end{equation}}
\newcommand{\ee}{\end{equation}}
\newcommand{\beqa}{\begin{eqnarray}}
\newcommand{\eeqa}{\end{eqnarray}}
\newcommand{\beqan}{\begin{eqnarray*}}
\newcommand{\eeqan}{\end{eqnarray*}}
\newcommand{\ba}{\begin{array}}
\newcommand{\ea}{\end{array}}

\newcommand{\hyp}{\mycal S}

\newcommand{\mcV}{{\mycal V}}

\newcommand{\warn}[1]
{\protect{\stepcounter{mnotecount}}$^{\mbox{\footnotesize
$
\bullet$\themnotecount}}$ \marginpar{
\raggedright\tiny\em
$\!\!\!\!\!\!\,\bullet$\themnotecount: {\bf Warning:} #1} }

\newcommand{\R}{\mathbb{R}}

\newcommand{\N}{\mathbb N}

\newcommand{\ptc}[1]{\mnote{{\bf ptc:}#1}}

\newcommand{\mcL}{{\mycal L}}

\newcommand{\beqar}{\begin{deqarr}}
\newcommand{\eeqar}{\end{deqarr}}

\newcommand{\beaa}{\begin{eqnarray*}}
\newcommand{\eeaa}{\end{eqnarray*}}

 \let\g=\gamma

\newcommand{\bethm}{\begin{theorem}}
\newcommand{\et}{\end{theorem}}
\newcommand{\bl}{\begin{Lemma}}

\newtheorem{Theorem} {\sc  Theorem\rm} [section]
\newtheorem{theorem} [Theorem] {\sc  Theorem\rm}

\newtheorem{cor} [Theorem] {\sc  Corollary\rm}

\newtheorem{Lemma} [Theorem] {\sc  Lemma\rm}
\newtheorem{Proposition} [Theorem] {\sc  Proposition\rm}

\newtheorem{definition}[Theorem]{\sc  Definition\rm}

\theoremstyle{Remark}

\theorembodyfont{\upshape}

\newtheorem{Remark}[Theorem]{\sc Remark\rm}
\newtheorem{remark}[Theorem]{\sc Remark\rm}

\theoremsymbol{\ensuremath{\diamondsuit}}
\newcommand{\fcoco}{\small}
\theorembodyfont{\fcoco}\theoremseparator{}\theoremindent0.5cm

\theoremstyle{nonumberplain}\theorembodyfont{\fcoco}
\theoremseparator{}\theoremindent0.5cm

\theoremstyle{definition}

\DeclareFontFamily{OT1}{rsfs}{}
\DeclareFontShape{OT1}{rsfs}{m}{n}{ <-7> rsfs5 <7-10> rsfs7 <10-> rsfs10}{}
\DeclareMathAlphabet{\mycal}{OT1}{rsfs}{m}{n}

{\catcode `\@=11 \global\let\AddToReset=\@addtoreset}
\AddToReset{equation}{section}

\renewcommand{\ptcn}[1]{} 
\renewcommand{\rnoblackholesc}[2]{}

\renewcommand{\bmcM}{{\,\,\,\,\widetilde{\!\!\!\!\mcM}}}

\renewcommand{\red}[1]{#1}
\renewcommand{\blue}[1]{#1}

\renewcommand{\ptcheck}[1]{}

\begin{document}
\title{%
Maximal hypersurfaces in asymptotically Anti-de Sitter spacetimes\protect\thanks{Preprint  UWThPh 2022-12}
}

\author{Piotr T.\ Chru\'{s}ciel\thanks{Email
\url{piotr.chrusciel@univie.ac.at}, URL \url{http://homepage.univie.ac.at/piotr.chrusciel}}
\\ University of Vienna, Faculty of Physics \\  Vienna, Austria
\\
\\
Gregory J. Galloway\thanks{Email
\url{galloway@math.miami.edu}, URL \url{http://math.miami.edu/\~galloway/}}
\\
University of Miami, Department of Mathematics  \\
Coral Gables, FL, USA}

\maketitle
\begin{abstract}
We prove uniqueness, existence, and regularity results for maximal hypersurfaces in  spacetimes with a conformal completion at timelike infinity and
asymptotically constant scalar curvature, as relevant for asymptotically AdS spacetimes.

This  work is dedicated to Yvonne Choquet-Bruhat on the occasion of her upcoming 99th birthday.
\end{abstract}
 
\setcounter{tocdepth}{1}
\tableofcontents

\definecolor{HP}{rgb}{1,0.09,0.58}

\section{Introduction}
It is a great pleasure to dedicate this work to Yvonne Choquet-Bruhat, who wrote pioneering papers~\cite{YvonneASNP,YvonneRMP} on this subject, and on many other ones.

Indeed, Yvonne has written several papers concerning existence and uniqueness properties of maximal (and CMC) hypersurfaces in spacetime manifolds. In \cite{YvonneASNP}  she states ``The existence of a maximal submanifold (with respect to area) is an important property for a space time, hyperbolic riemannian manifold satisfying Einstein equations."  She goes on to describe the importance of this for solving the Einstein constraint equations, and also the relevance of this (at the time) for proving positivity of mass.

Meanwhile the existence of maximal hypersurfaces has been found to be useful for other purposes as well.  Of particular relevance here is a classical result of Bartnik~\cite{bartnik:variational} that establishes, under certain conditions, the existence of maximal hypersurfaces in asymptotically flat spacetimes. This and related results have been useful for a number of purposes in the study of such spacetimes.  Given the role of asymptotically Anti-de Sitter spacetimes in theoretical physics (e.g. via the AdS/CFT correspondence), one is naturally led to consider maximal hypersurfaces in this setting, where now the notion of renormalised volume also becomes relevant.

The question then arises of existence of maximal hypersurfaces
in $(n+1)$-dimensional, $n\ge 2$, asymptotically locally hyperbolic spacetimes, with controlled asymptotic behaviour at the conformal boundary at infinity so that the renormalised volume is defined.
The simplest case where an affirmative answer can be given is in a perturbative setting, when the metric, or the asymptotic data, or both, are perturbed away from a maximal slicing, and when an energy condition is satisfied. We prove this in  Theorem~\ref{T4X21.1} below, generalising a related result of \cite{ShiMax}.

However,
simple examples show that some global regularity conditions need to be satisfied by the spacetime for existence in general.
Here two conditions arise naturally: that of existence of barriers, or that of compactness of the domain of dependence of a fiducial Cauchy surface. Under these two conditions we show solvability of an asymptotic Dirichlet problem for the  maximal hypersurface equation; see Theorems~\ref{T11IX21.1a} and \ref{T11IX21.1}. This generalises previous results of \cite{Akutagawa,BonsanteSchlenkerTeichmuller}, established under restrictive conditions on the dimension or on the class of spacetimes considered.

As such, these last two theorems do not guarantee that a well defined notion of renormalised volume exists
without further conditions; see for instance   Theorem~\ref{T28VII22.1} where both ``good barriers'' and  timelike convergence is assumed.
This is related to the question of the behaviour of the maximal hypersurface at the conformal boundary at infinity, which turns out to be delicate. Indeed, while maximal hypersurfaces are typically spacelike and smooth in the interior, they might become asymptotically null when the conformal boundary is approached.  We show that maximal hypersurfaces which are uniformly spacelike at infinity, so that the last possibility does not occur, are uniquely defined by their asymptotic Dirichlet data, meet the conformal boundary orthogonally, and have a full polyhomogeneous expansion at infinity, in particular a well defined renormalised volume; see Theorems~\ref{T3IV22.1}  and \ref{T3IV22.2} for precise statements.

Our analysis shows that a possible obstruction to regularity of maximal hypersurfaces at the conformal boundary at infinity is that the hypersurface becomes asymptotically null. The key remaining question is
to show that asymptotically null maximal hypersurfaces do not exist, or to construct a counterexample.

\section{Preliminaries}

Let $(\mcM,\fourg)$ be an $(n+1)$-dimensional spacetime with a time function $t$. Consider a coordinate patch $(t,x^i)$, in which we write the metric as
\begin{equation}\label{28IV21.1a}
  \fourg = - \alpha^2 dt^2  + g_{ij}(dx^i +\beta^i dt)(dx^j+\beta^j dt)
  \,.
\end{equation}
Hence $\det \fourg = - \alpha^2 \det g$ and
%
\begin{equation}\label{28IV21.1b}
  \fourg^\sharp = -\alpha^{-2} \partial_t^2
   + 2\alpha^{-2}  \beta^i   \partial_t\partial_i + (g^{ij}-   \alpha^{-2} \beta^i \beta^j) \partial_i \partial_j
  \,,
\end{equation}
where $g^{ij}$ denotes the matrix inverse to $g_{ij}$.

\subsection{The mean curvature operator}
 \label{ss9V21.1}

In the coordinate system above, let a spacelike hypersurface $\hyp$ be given by the equation $t=u(x^i)$. Let $N$ be  the future-directed unit normal to $\hyp$ and let $\{e_\mu\}_{\mu=0}^n$ be an ON basis of  $T\mcM $ defined in a neighborhood of $\hyp$ such that $e_0=N$ along $\hyp$,
thus $\fourg(e_\mu,e_\nu) = \mathrm{diag}(-1,1,\ldots,1)$.
The mean curvature of $H$ is given by the equation
\begin{equation}\label{28IV21.2}
  H =
   \sum_{i=1}^n \fourg(e_i,\nabla_{e_i}N)=
      \underbrace{\fourg(N,\nabla_N N)}_{= \frac 12 \nabla_N \fourg(N,N) = 0} + \fourg^{\mu\nu} \fourg(e_\mu,\nabla_{e_\nu}N)
    = \nabla_\mu N^\mu
   \,,
\end{equation}
where
 $\nabla$   denotes the Levi-Civita derivative of the metric $\fourg$.
We have
\ptcheck{29IV21}
\begin{eqnarray}
 N^\flat
 &
  \displaystyle =
 & - \frac{dt -du}{\sqrt{|\fourg^{tt} - 2 \fourg^{ti} \partial_i u + \fourg^{ij} \partial_i u \partial_ju |}}
  \equiv
    -\frac{ \alpha (dt - \partial_i u \,dx^i)}{\sqrt{(1  +  \beta^i  \partial_i u)^2 - \alpha^2 g^{ij}  \partial_i u \partial_ju }}
   \nonumber
\\
&
\Longleftrightarrow
&
 \nonumber
\\
  N
 &
 \displaystyle =
   &- \frac{\fourg^{t\mu} -\fourg^{i\mu}\partial_iu}{\sqrt{\fourg^{tt} - 2 \fourg^{ti} \partial_i u + \fourg^{ij} \partial_i u \partial_ju }}\partial_\mu
    \nonumber
\\
 &  \equiv &
    \frac{  \partial_t  - \beta^i \partial_i + \beta^i \partial_i u \partial_t
      +
      \alpha^2  (g^{ij}-   \alpha^{-2} \beta^i \beta^j) \partial_i u \partial_j }{ \alpha \sqrt{(1  +  \beta^i  \partial_i u)^2 - \alpha^2 g^{ij}  \partial_i u \partial_ju }}
\\
 &  =|_{\beta = 0}  &
    \frac{  \partial_t
      +
      \alpha^2   g^{ij}  \partial_i u \partial_j }{ \alpha \sqrt{1- \alpha^2 g^{ij}  \partial_i u \partial_ju }}
      \equiv
       \frac{  \partial_t
      +
      \alpha^2  D^i u \partial_i }{ \alpha \sqrt{1- \alpha^2 |Du|_g^2 }}
   \,,\label{28IV21.3}
\end{eqnarray}
with $D$ denoting the Levi-Civita derivative of the metric $g:= g_{ij}dx^i dx^j$. This gives
\begin{eqnarray}
  H
   & = &
    \nabla_\mu N^\mu =
   \frac{1}{\sqrt{|\det \fourg|}} \partial_\mu
   \left(\sqrt{|\det \fourg| }N^\mu \right)
    \nonumber
\\
 & = &
   \frac{1}{\alpha \sqrt{ \det g}} \partial_t
   \left(\alpha \sqrt{ \det g }N^t \right)
   +
   \frac{1}{\alpha \sqrt{ \det g}} \partial_i
   \left(\alpha \sqrt{ \det g }N^i \right)
    \nonumber
\\
 & = &
   \frac{1}{\alpha \sqrt{ \det g}} \partial_t
   \left(\alpha \sqrt{ \det g }N^t \right)
   +
   \frac{1}{\alpha } D_i
   \left(\alpha  N^i \right)
 \,.
   \label{9V21.1}
\end{eqnarray}
It is convenient to define
\begin{eqnarray}
  T & : = &  - \alpha  \nabla t \equiv \alpha^{-1} ( \partial_t  - \beta^i \partial_i )
  \,,
\\
   \nu
    &:= &
      -\fourg(T,N)
   \equiv
    \frac{  1  + \beta^i \partial_iu}
    {  \sqrt{(1  +  \beta^i  \partial_i u)^2 - \alpha^2 g^{ij}  \partial_i u \partial_ju }}
     \label{3IV22.22}
\\
    & = &\!\!\!\!\!|_{\beta =0}
     \,\,\,\,
    \frac{  1  }
    {  \sqrt{1 - \alpha^2 |Du|_g^2 }}
   \,,
    \label{28IV21.4}
\end{eqnarray}
thus $T$ is the field of unit normals to the level sets of $t$.
The function $\nu$ controls the slope of the graph of $u$: indeed, $1 - \alpha^2 |Du|_g^2$ vanishes at $p$ if and only if the graph of $u$ is null at $p$. So a graph of a continuously differentiable function $u$  is spacelike if and only if $\nu$ is uniformly bounded on every compact subset of the graph.

As emphasised by Bartnik~\cite{Bartnik84,bartnik:variational}, the function $\nu$ plays a key role when studying maximal hypersurfaces, and it will likewise play a key role here. Following Bartnik, we will refer to $\nu$ as the \emph{tilt function}.
In physics $\nu$ is the
relative $\gamma$-factor of
the local inertial frames associated with $T$ and $N$.

We note that $N^t=\alpha^{-1}\nu$, so that we can rewrite \eqref{9V21.1} as
\begin{eqnarray}
 H
   & =&
   \frac{1}{\alpha \sqrt{ \det g}} \partial_t
   \left( \sqrt{ \det g } \, \nu \right)
   \nonumber
\\
 &&
   +
   \frac{1}{\alpha } D_i
   \left(
   \frac{
       - \beta^i
      +
      \alpha^{\red{2}}  (g^{ij}-   \alpha^{-2} \beta^i \beta^j) \partial_j u  }
      {  \sqrt{(1  +  \beta^i  \partial_i u)^2 - \alpha^2 g^{ij}  \partial_i u \partial_ju }}
       \right)
        \label{11X21.2}
\\
 &  =|_{\beta =0}   &
  \frac{1}{\alpha \sqrt{ \det g}} \partial_t
   \left( \frac{\sqrt{ \det g }}{ \sqrt{1 - \alpha^2 |Du|_g^2 }} \right)
   +
   \frac{1}{\alpha } D_i
   \left(\frac{
      \alpha^{\red 2} D^i u }{ \sqrt{1 - \alpha^2 |Du|_g^2 }}  \right)
      \!\!
 \,. \phantom{xxxx}
   \label{9V21.2}
\end{eqnarray}

In order to make contact with the notation in~\cite{Bartnik84,Akutagawa}
we let $\divM$ be the divergence on $\hyp$,
\begin{equation}\label{28IV21.5}
  \divM Y := \sum_{i=1}^n \fourg(e_i,\nabla_{e_i} Y)
  \,,
\end{equation}
where $e_i$ is any $ON$-frame spanning $T\hyp$. If $Y$ is tangent to $\hyp$ this is simply the divergence of $Y$ with respect to the metric induced on $\hyp$. This leads to the following rewriting of
\eqref{28IV21.2} as in \cite{Bartnik84}:
\begin{equation}\label{28IV21.2a}
  H = \nu^{-1}
    \Big(
     \divM \underbrace{(\nu N - T)}_{\alpha \gradM u} + \divM   T
     \Big)
\end{equation}
(keeping in mind that $ \divM  (\nu N)= \nu \,\divM   N $),
where the vector field
$$   \gradM u := \alpha^{-1}(\nu N - T)
\,,
$$
which vanishes if and only if $\nabla u $ vanishes,
carries geometric information about the projection of $\nabla u$  to $T\hyp$. Here, and elsewhere in this work, $u$ is understood as a function on spacetime satisfying $\partial_t u = 0$.

\subsection{Asymptotically locally hyperbolic metrics}
 \label{ss9V21.2}

Suppose that $\red{(\mcM,\fourg)}$
has a smooth conformal completion $\red{(\bmcM,\widetilde \fourg)}$
at timelike conformal infinity $\tScri:=\bmcM\setminus \mcM$,
with the scalar curvature of $\fourg$ tending to a constant as $\tScri$ is approached.
Such metrics will be called \emph{asymptotically locally hyperbolic} (ALH).
In what follows we will always assume that partial Cauchy surfaces of $\tScri$ are compact.

Let
$$
 (x^\mu)\equiv (x^0,x^i)\equiv (t,x^i)\equiv (x^0,x^A,x) \equiv (x^a,x)
  \,,
$$
where $a\in\{0,1,\ldots,n-1\}$, be Fefferman-Graham-type coordinates in which
$$\tScri = \{x=0\}
\,,
$$
and in which $\fourg= x^{-2} \tilde \fourg $ takes the form
\begin{eqnarray}
  \fourg
   &=& x^{-2} (dx^2 + \gamma_{ab}dx^a dx^b )
    \nonumber
\\
    &=&  x^{-2} \left(
     dx^2 + \gamma_{tt}dt^2 +2   \gamma_{tA}dx^A dt
    +   \gamma_{AB}dx^A dx^B
     \right)
    \nonumber
\\
    &=:&  x^{-2} \left(dx^2
     - \hat \alpha^2 dt^2  + h_{AB}(dx^A +\hat \beta^A dt)(dx^B+\hat \beta^B dt)
      \right)
     \,.
     \label{9V21.6}
\end{eqnarray}

Comparing with \eqref{28IV21.1a} we find
\begin{eqnarray}
&
  \alpha =x^{-1}\hat \alpha
   \,,  \quad
   \fourg_{AB} =g_{AB} = x^{-2} h_{AB}
   \,,
   \quad
  \sqrt{|\det \fourg|}  =x^{-(n+1)} \hat \alpha \sqrt{\det h}
   \,,
  &
   \label{9V21.7a}
\\
 &
  \sqrt{|\det g|}  =x^{-n}  \sqrt{\det h}
   \,,
   \quad
  \beta^x \equiv 0
   \,, \quad
  \fourg_{tA} = g_{AB} \beta^B =x^{-2} h_{AB} \hat \beta^B
  \ \Rightarrow \
    \beta^A = \hat \beta^A
   \,.
    &
   \phantom{xxxx}
   \label{9V21.7b}
\end{eqnarray}
  For further reference we note
\begin{equation}\label{28IV21.1btd}
  \fourg^\sharp = x^2
  \Big(
  \partial_x^2
   -\hat \alpha^{-2} \partial_t^2
   + 2\hat \alpha^{-2}  \hat \beta^A   \partial_t\partial_A + (h^{AB}-   \hat \alpha^{-2} \hat \beta^A \hat \beta^B) \partial_A \partial_B
    \Big)
  \,,
\end{equation}
where $h^{AB}$ is the inverse metric of $h_{AB}$.

Assume   that $\hat \beta^ {A} \equiv 0$. From \eqref{9V21.2} we find
%
\begin{eqnarray}
 H
   & =&
  \frac{x }{ \hat \alpha  \sqrt{ \det h}} \partial_t
   \left( \frac{  \sqrt{ \det h }}
    { \sqrt{1 - \hat  \alpha^2 ((\partial_xu)^2+|\mcD u|_h^2) }} \right)
   +
  \frac{x^{ } }{\hat \alpha  }  \mcD_A
   \left(\frac{ \hat \alpha^{\red 2}  \mcD^A u }{ \sqrt{1 - \hat  \alpha^2 ((\partial_xu)^2+|\mcD u|_h^2) }} \right)
 \nonumber
\\
 &&
      +
  \frac{x^{n+1} }{\hat \alpha \sqrt{ \det h}} \partial_x
   \left( \frac{ x^{ -n}\hat \alpha^{\red 2} \sqrt{ \det h }  \partial_x u
   }{ \sqrt{1 - \hat  \alpha^2 ((\partial_xu)^2+|\mcD u|_h^2) }} \right)
 \,,
   \label{9V21.7c}
\end{eqnarray}
where $\mcD$ denotes the Levi-Civita covariant derivative associated with (the $(t,x)$-dependent family of metrics) $h\equiv h_{AB}dx^Adx^B$.   In \eqref{9V21.7c} all the metric functions are evaluated on the graph $(t=u(x^i), x^i)$,
but expressions such as  $\partial_i \hat \alpha$ are  standard partial derivatives,
and \emph{not} $\partial_i u\partial_t \hat \alpha + \partial_i \hat \alpha$.

\section{Boundary behaviour for $C^2$ up-to-boundary maximal hypersurfaces}
  \label{s9V21.4}

We wish to use the results in \cite{Akutagawa} and  \cite[Chapter~5]{AndChDiss}  to establish the asymptotic behaviour at the conformal boundary at infinity of a class of solutions of the equation $H=H_1$, where $H_1$ is a smooth function on spacetime.

\subsection{Asymptotic expansions}
 \label{ss16IX21.1}

Some insight into the behaviour of the solutions is obtained by constructing approximate solutions of the equation at hand using asymptotic expansions. It will be seen in Section~\ref{ss11X21.1} that this is in any case useful when studying the asymptotic behaviour of the solutions near the conformal boundary at infinity.

As already mentioned, we consider the equation
\begin{equation}\label{17IX21.12}
 H[u]=H_1
  \,,
\end{equation}
where $H_1$ is a smooth function on $\mcM$. We are interested in solutions which are graphs $t=u(x^i)$, in coordinates as in \eqref{9V21.6}.

We start with the case
\begin{equation}\label{16IX21.11}
  H_1 = 
  \ozero{H}_1 + O(x)
\qquad
\Longleftrightarrow
\qquad
 \ozero{H}_1 = H_1\big|_{x=0}
  \,,
\end{equation}
where $\ozero{H}_1 :\red{\tScri} \mapsto \R $ is not {necessarily} zero.
We write the function $\hat \alpha$ appearing in \eqref{9V21.7a} as
\begin{equation}\label{11X21.1}
  \hat\alpha = \red{\mathring{\hat \alpha}} + O(x)
   \,,
   \quad
   \mbox{with}\ 0< \red{\mathring{\hat \alpha}}:  \red{\tScri} \mapsto \R
   \,.
\end{equation}
When $\hat \beta^i\big|_{x=0} =0$
 one easily checks
(compare \eqref{9V21.7c})
 that we will have
\begin{equation}\label{17IX21.12a}
 H[x \oone
 u + o(x)]= \ozero{H}_1 + O(x)
\,,
\end{equation}
where the $o(x)$ and $O(x)$=terms in the equation behave under differentiation in  the obvious way,
if
\begin{equation}\label{17IX21.13}
   \frac{\red{\mathring{\hat \alpha}} \oone u}{\sqrt{1-(\red{\mathring{\hat \alpha}} \oone u)^2}} = -\frac{\ozero{H}_1 }n
   \quad
   \Longleftrightarrow
   \quad
   \oone u = -\frac{   \ozero{H}_1 }{\red{\mathring{\hat \alpha}} \sqrt{n^2 + (\ozero{H}_1)^2}}
  \,,
\end{equation}
and note that  $|\red{\mathring{\hat \alpha}} \oone u| < 1$ for all $\ozero{H}_1$. Inspection of \eqref{11X21.2} shows that these formulae remain valid  when  $\hat \beta^i\big|_{x=0} \ne 0$.

For maximal hypersurfaces we have $\ozero{H}_1=0$, so that an immediate corollary of \eqref{17IX21.13} is:

\begin{Proposition}
  \label{P4IV22.1}
Let $\hyp$ be a maximal hypersurface  in $\mcM$ which is $C^2$-up-to-boundary, and spacelike up-to-boundary in the conformally rescaled metric. Then $\hyp$ meets the conformal boundary of $\mcM$ orthogonally.
\end{Proposition}

It turns out that
 \eqref{17IX21.13} will play an important role in the analysis of the equation of main interest to us, namely $H[u]=0$, where $\ozero{H}_1 = 0$.
     The analysis of the mean-curvature equation with   $\ozero{H}_1 \ne 0$   introduces  complications which are irrelevant for the purpose of studying maximal hypersurfaces.
    While it is likely that one can develop a regularity theory similar to the one in Section~\ref{ss11X21.1} below in the more general case, for the sake of simplicity  in several results in this work, including in what follows in the current section, we will assume that
    $$
    \ozero{H}_1 \equiv 0
    \,.
    $$

Suppose then that all the metric functions in \eqref{9V21.6} are smooth up to the boundary $\{x=0\}$. It follows from  \eqref{17IX21.12a}
that the mean curvature, say $H_0$,
 of the slice $\hyp=\{t=0\}$ is $O(x)$. Let us therefore assume   that there exists an integer $\ell \ge 1$ so that  $H_0$
  has the  Taylor expansion
\begin{equation}\label{18IX21.1a}
  H_0 =  \oell{H}_0 x^\ell  +  \oellp{H}\!\!\!_0 x^{\ell+1} + \ldots
  \,.
\end{equation}
Let us further suppose that $H_1$ has a similar expansion
\begin{equation}\label{18IX21.1b}
  H_1 = \oell{H}_1 x^\ell +  \oellp{H}\!\!\!_1 x^{\ell+1} + \ldots
  \,.
\end{equation}
In the simpler case $\hat \beta ^i\equiv 0$,
matching powers of $x$ in \eqref{9V21.7c}  one finds approximate solutions of the form
\begin{equation}\label{18IX21.2}
  u  =
  \left\{
    \begin{array}{ll}
        \oellp{u}  x^{\ell+1} +  \oellpp{u}  x^{\ell+2} + \ldots, &
 \hbox{$\ell \ne n-1,\,n $ ;}
\\
      \oellp{u}  x^{\ell+1}  +  \onlog {u}  x^{n+1} \log x +  \onp {u}  x^{n+1} + \ldots, &
 \hbox{$\ell=n-1 $,}
\\
      \onlog{u}  x^{n+1} \log x +  \onp {u}  x^{n+1} + \ldots, & \hbox{$\ell=n $ ,}
    \end{array}
  \right.
\end{equation}
with smooth expansion coefficients $\oellp u$, $\oellpp u$, $\onlog u$,  etc.   For example, we have
\begin{equation}\label{18IX21.2a}
    \begin{array}{ll}
        \displaystyle
\oellp{u} =
- \frac{\oell{H}_1-\oell{H}_0}{ (n-\ell)(\ell+1)\red{\mathring{\hat \alpha}}  }
, &
 \hbox{$\ell \ne n $ ;}
\\
        \displaystyle
   \onlog {u} =
 \frac{\on{H}_1-\on{H}_0}{(n+1 )\red{\mathring{\hat \alpha}}  }, &
 \hbox{$\ell = n $ .}
    \end{array}
\end{equation}
One checks that this scheme remains true for general smooth $\hat \beta$'s, keeping in mind that $\beta^x\equiv 0$ in Fefferman-Graham coordinates.

Recall that a function is called \emph{polyhomogeneous} if it admits,  for small $x$, an asymptotic expansion in terms of the functions $x^i\ln^j x$, with smooth expansion coefficients. (See e.g.\ \cite{AndChDiss} for a  detailed presentation.)
When $\ell\le n$ the approximate solution \eqref{18IX21.2} might pick-up a logarithmic term of order $x^{n+1}\log x$, and will continue with higher powers of $x$ as a polyhomogeneous expansion. When $\ell > n+1$ the approximate solution will have a complete asymptotic expansion in powers of $x$, without log terms; in this case the \emph{approximate solution} can be chosen to start at $x^{\ell+1}$, but note that an associated \emph{solution} will generically have an expansion with  log terms starting with the power $x^{n+1}$.

\subsection{The linearised operator}

In what follows  we will need  the linearisation $H'|_u$ of the operator $H$ at a function $u$ which is differentiable up-to-boundary. Assuming $\hat \beta^A\equiv 0$  we find from \eqref{9V21.7c}, using hopefully obvious notation,
\begin{eqnarray}
\lefteqn{
 H'|_u [v]
   \equiv  \frac{dH }{du}[v] =
  \frac{d}{du}
  \Bigg[\frac{x}{\hat \alpha  \sqrt{ \det h}} \partial_t
   \Bigg( \frac{ \sqrt{ \det h }}
    { \sqrt{1 - \hat  \alpha^2 ((\partial_xu)^2+|\mcD u|_h^2) }} \Bigg)
    }
   &&
   \nonumber
\\
 &&
   +
  \frac{x }{\hat \alpha  }  \mcD_A
   \Bigg(\frac{ \hat \alpha^2  \mcD^A u }{ \sqrt{1 - \hat  \alpha^2 ((\partial_xu)^2+|\mcD u|_h^2) }} \Bigg)
   \Bigg] [v]
 \nonumber
\\
 &&
      +
   {x^{n+1} }\partial_x
   \left( \frac{ x^{-n}\hat \alpha^{\red 2} \sqrt{ \det h }  \partial_x u }{ \sqrt{1 - \hat  \alpha^2 ((\partial_xu)^2+|\mcD u|_h^2) }} \right)
    \frac{\partial}{\partial t}\left(\frac{1}{\hat \alpha  \sqrt{ \det h}} \right) v
 \nonumber
\\
 &&
   +
  \frac{x^{n+1} }{\hat \alpha  \sqrt{ \det h}} \partial_x
   \Bigg(
    \frac{ x^{-n}\hat \alpha^{\red 2} \sqrt{ \det h }  \, \partial_x v }{ \sqrt{1 - \hat  \alpha^2 ((\partial_xu)^2+|\mcD u|_h^2) }}  +
    \frac{ x^{-n}\partial_t(\hat \alpha^{\red 2} \sqrt{ \det h } ) \partial_x u }{ \sqrt{1 - \hat  \alpha^2 ((\partial_xu)^2+|\mcD u|_h^2) }}   v
 \nonumber
\\
 &&
   +
    \frac{  x^{-n}\hat \alpha^{\red 2} \sqrt{ \det h }  \partial_x u
    \Big(
     \hat \alpha \partial_t \hat \alpha
      \big((\partial_xu)^2+|\mcD u|_h^2 \big) v
    +\hat  \alpha^2 ( \partial_xu \, \partial_x v +h(\mcD u, \mcD v)
    + \frac 12
     (\partial_t h^{AB}\mcD_A u \mcD_Bu) v
    \big)\Big)}{(1 - \hat  \alpha^2 ((\partial_xu)^2+|\mcD u|_h^2  )^{\frac{3}{2}}} \Bigg)
 \,,
 \nonumber
\\
 &&
   \label{8VI21.1}
\end{eqnarray}
where the $d/du[...][v]$ term, which arises from the first line in \eqref{9V21.7c}, has not been made explicit  as it is not very englightening in the generality.
Equation~\eqref{8VI21.1} simplifies somewhat   at a constant function $u$, say $u=0$:
\begin{eqnarray}
 H'|_{u=0} [v]
 &=&
  \partial_t
  \Bigg(\frac{x}{\hat \alpha  \sqrt{ \det h}} \partial_t
   \left(  { \sqrt{ \det h }} \right) \Bigg) v
   +
  \frac{x }{\hat \alpha  }  \mcD_A
   \left(  \hat \alpha^2  \mcD^A v \right)
 \nonumber
\\
 &&
   +
  \frac{x^{n+1} }{\hat \alpha  \sqrt{ \det h}} \partial_x
   \Big (
      x^{-n}\hat \alpha^{\red 2} \sqrt{ \det h }  \, \partial_x v
      \Big)
 \,.
   \label{9VI21.2}
\end{eqnarray}
%


Keeping in mind that we have assumed conformal smoothness of the metric, after multiplication by $x$ the operator \eqref{9VI21.2}   becomes a linear partial differential operator of Fuchsian type.
Its \emph{indicial exponents} $\sigma_-<\sigma_+$ can be calculated from the equation
\begin{equation}\label{8VI21.2}
 x H'|_{u=0} [x^{\sigma_\pm} ] = o(x^{\sigma_\pm})
  \,.
\end{equation}
We have
\begin{equation}\label{8VI21.3}
  x H'|_{u=0} [x^\sigma  ] = \sigma(\sigma-n-1) \hat \alpha x^\sigma + o(x^\sigma)
  \,,
\end{equation}
hence
\begin{equation}\label{8VI21.4}
  \sigma_-=0
  \,,
  \quad
   \sigma_+ = n +1
   \,.
\end{equation}

The above analysis still applies if $u$ is, say, a polyhomogeneous function satisfying $u=o(x)$ for small $x$.
 It remains true for the linearisation of the full operator $H$ with non-vanishing shift vector $\beta^i$ if moreover $\hat \beta^A$ vanishes at $x=0$, which can be typically achieved by a change of coordinates   near the conformal boundary.
\subsection{Polyhomogeneous solutions}
\label{ss11X21.1}

The results in the last sections lead to the following:

\begin{Theorem}
  \label{T19IX21.1}
 In the setting of Section~\ref{ss9V21.2}, let $H_1\in C^\infty(\omcM)$ satisfy $H_1\big|_{x=0}=0$.  In the coordinates of \eqref{9V21.6} consider a spacelike graph  $t=u(x^i)$ over $\ovhyp:=\{t=0\}$ solving the equation
  \begin{equation}
   H[u]=H_1
   \,,
   \label{21IX21.1}
   \end{equation}
such that
\begin{equation}\label{4IV22.1}
  \mbox{$u=O(x^{1+\epsilon})$ and   $|D u|_g=O(x^{1+\epsilon})$ for some $\epsilon>0$.}
\end{equation}
Then $u$ is $C^n(\ovhyp)$ and polyhomogenous.
\qed
\end{Theorem}

\begin{Remark}
  \label{R19IX21.1}
  In Fefferman-Graham coordinates the condition $|D u|_g=O(x^{1+\epsilon})$ implies $ \partial_x u =O(x^{ \epsilon})$, and is equivalent to the last condition for polyhomogeneous or smooth functions.
  It will be satisfied if $u$ is $C^2$ up to boundary with vanishing derivative there.
  The graphing function $u$ will be asymptotic to the approximate solution  $u_0$ of Section~\ref{ss16IX21.1} to infinite order, which implies that its differentiability properties at $\red{\tScri}$ will be identical to those of the approximate solutions described there.
  \qed
  %
\end{Remark}

\begin{Remark}
  \label{R19IX21.2}
  If both the conformally rescaled metric and $H_1$ are in $C^k(\ovmcM)$ and polyhomogeneous, then $u$ will be
   $C^{\min(k+1,n)}(\ovhyp)$ and polyhomogeneous, with the coefficients of the polyhomogeneous expansion determined by those of $H_1$ and of the metric.
  \qed
  %
\end{Remark}

\proof
We check that the results in~\cite[Chapter 5]{AndChDiss} apply. For this we set
\begin{equation}\label{11VI21.1}
  u = x \hat u
  \,,
\end{equation}
and, as in \cite[Equation~(5.1.1)]{AndChDiss}, we rewrite the equation $H=H_1$
 as an equation $F[\hat u, x\partial_i \hat u,x^2 \partial_i \partial_j \hat u]= 0$. We need to check the hypotheses of~\cite[Chapter 5]{AndChDiss}. The first condition is  the existence of polyhomogeneous approximate solutions, say $u_0$ of the equation. This has already been established in Section~\ref{ss16IX21.1}.
The second condition is the existence of a ``regularity interval'' for the operator obtained by linearising the equation at $u_0$. In our case this operator is the Laplace operator perturbed by lower order terms, and the existence of the regularity interval follows from \cite[Section~7.2]{AndChDiss}; cf.\ Remark (i) on p.\ 77 there. The result follows now from \cite[Proposition~5.1.5]{AndChDiss}.
\qed

\section{Akutagawa's derivative estimate}
 \label{s19III22.1}

In this section we provide a generalisation of an estimate of Akutagawa~\cite{Akutagawa}, as useful for the problem at hand.

We introduce the Riemannian metric
\begin{eqnarray}
  \backriemg
    &:=&  \fourg+ 2 T^\flat \otimes T^\flat
    \nonumber
\\
 &= &
    x^{-2} \left(dx^2
     + \hat \alpha^2 dt^2  + h_{AB}(dx^A +\hat \beta^A dt)(dx^B+\hat \beta^B dt)
      \right)
     \,.
     \label{9V21.6a}
\end{eqnarray}
The symbol $|\cdot|_{\backriemg}$ denotes  the norm of a tensor with respect to this metric.

The following result,
essentially due to
Akutagawa \cite[Proposition~1]{Akutagawa} (see Remark~\ref{R10VI22.1} below),
provides a $C^1$ bound for the graphing function when a height bound and the mean curvature of the graph are known. The key difference between \cite[Proposition~1]{Akutagawa} and Bartnik's \cite[Theorem~3.1]{Bartnik84} is that
Bartnik's hypothesis  of boundedness of the lapse function $ \alpha $, which is unbounded in our context, is replaced by Akutagawa with the \emph{a priori} condition
\eqref{20III22.2b}  below on $u$. We emphasise that the further details of the global geometry of $(\mcM,g)$ are irrelevant here, in particular existence of a conformal completion of $\mcM$ is \emph{not} assumed.

\begin{Theorem}
  \label{T19III22.1}
  Let $H$ denote the mean curvature of the graph of $u$ over the closure $\overline \Omega$ of a conditionally compact domain $\Omega \subset \{t=0\}$,
   and let $\Ric$ be the Ricci tensor of $\fourg$. Suppose that there exists a  constant  $C$  such that  along the graph we have
\begin{eqnarray}
 &
   \displaystyle
   |\Ric |_{\backriemg} +  |  \mcL_T \fourg|_{\backriemg}
   +  |\nabla \mcL_T \fourg|_{\backriemg}
   +  |\alpha^{-1}   \nabla \alpha|_{\backriemg}
   +  |\alpha^{-1} \nabla \nabla \alpha|_{\backriemg}
   \le C
   \,,
   &
    \label{20III21.1-1}
\\
 &
   \displaystyle
    | H|
   +|\nabla H |_{\backriemg}
   \le C
   \,.
   &
    \label{20III21.1}
\end{eqnarray}
If $\partial \Omega\ne \emptyset$, suppose moreover that $u|_{\partial \Omega } =0$ and that the mean  curvature $H_{\partial \Omega}$ vector of $\partial \Omega$ satisfies
\begin{equation}\label{27III22.1b}
  | H_{\partial \Omega}|_{\backriemg} \le C
  \,.
\end{equation}
If there exist positive constants $\delta$ and $\hat C$  such that
\begin{eqnarray}
 &
   \displaystyle
   |u | \le \hat C \alpha ^{1+\delta}
   \,,
    \label{20III22.2b}
   &
\end{eqnarray}
then there exists a constant $\tilde C(C,\hat C, \delta)$ such that
\begin{eqnarray}
 &
   \displaystyle
   \nu \le \tilde C
   \,.
    \label{27III22.2b}
   &
\end{eqnarray}
\qed
\end{Theorem}

\begin{remark}
 \label{R10VI22.1}
Akutagawa explicitly assumes space-time dimension four, which is irrelevant except for changing some  numerical coefficients in his argument.
Inspection of \cite[Proposition~1]{Akutagawa} shows that the condition of boundedness of  $ |\Ric(V,V) |$ in \cite[Equation~(3.3)]{Akutagawa} can be replaced by  boundedness of $|\Ric |_{\backriemg}$ in the proof there.
This  is better suited to our purposes,  as the assumption of boundedness of  $ \Ric(V,V) $  is a condition involving both the graphing function $u$ and the metric, while $|\Ric |_{\backriemg}$ is independent of $u$.
Finally, both Bartnik's and Akutagawa's hypotheses on
    the second fundamental form $ |\mathring A  |_{\backriemg}$ of the level sets of $t$ with respect to the Lorentzian metric $\fourg$ are redundant, in that they already follow from the bound on $|  \mcL_T \fourg|_{\backriemg}$.
    \qed
\end{remark}

\section{Local barriers}
 \label{s27III22.1}

In this section we construct local barriers near the conformal boundary at infinity, as needed for controlling the solutions there.

When $u$ depends only upon $x$, we find
\begin{eqnarray}
   \nu
    &= &
    \frac{  1  }
    {  \sqrt{ 1   -  \hat \alpha^2  (\partial_x u)^2 }}
   \,,
    \label{28IV21.4td}
\end{eqnarray}
and
\eqref{11X21.2} becomes
\begin{eqnarray}
 H
   & =&
  \frac{x^{n+1 } }{\hat \alpha \sqrt{ \det h}}
   \partial_x
   \left( \frac{ x^{-n}\hat \alpha^{\red 2} \sqrt{ \det h } \partial_x u }{ \sqrt{1 - \hat  \alpha^2 (\partial_xu)^2 }} \right)
   +
   \frac{x }{\hat \alpha \sqrt{ \det h}} \partial_t
   \left(  \frac{  \sqrt{ \det h} }
    {  \sqrt{1   -  \hat \alpha^2  (\partial_x u)^2 }} \right)
    \nonumber
\\
   &  &
   -
  \frac{x  }{\hat \alpha \sqrt{ \det h}}
   \partial_A
   \left( \frac{ \sqrt{ \det h } \hat \beta^A  }{ \sqrt{1 - \hat  \alpha^2 (\partial_xu)^2 }} \right)
   \nonumber
\\
   & =&
  \underbrace{
    \frac{x^{n+1 } }{\hat \alpha \sqrt{ \det h}}
   \partial_x
   \left( \frac{ x^{-n}\hat \alpha^{\red 2} \sqrt{ \det h } \partial_x u }{ \sqrt{1 - \hat  \alpha^2 (\partial_xu)^2 }} \right)
   }_{(1)}
   +
   \underbrace{
      \frac{x \big(  \partial_t \hat \alpha  -  \hat \beta^A \partial_A \hat \alpha \big) (\partial_x u)^2 }
    {
    ( 1   -  \hat \alpha^2  (\partial_x u)^2
    )^{3/2}}
    }_{(2)}
    \nonumber
\\
   &  &
   +
   \underbrace{
   \frac{x\hat \alpha ^{-1}
    \big( \partial_t(\ln \sqrt{ \det h}) - \mcD_A \hat \beta^A
     \big)
       }
    {  \sqrt{1   -  \hat \alpha^2  (\partial_x u)^2 }}
    }_{(3)}
  \,,
   \nonumber
\\
(1)
 & = &
 \nu \Big[
 - \Big(n\hat \alpha- x \big(2
   \partial_x \hat \alpha +  \hat \alpha \partial_x(\ln \sqrt{ \det h})
   \big)
   \Big)
   \partial_x u
  \nonumber
\\
 &  &
   + x
    \Big((1
 + \hat \alpha^2
  \nu^2(\partial_xu)^2 ) \hat \alpha  \partial^2_x u
  +
 \hat \alpha^2  \nu^2 \partial_x \hat \alpha (\partial_x u )^3
   \Big)
    \Big]
 \,. \phantom{xxxx}
   \label{9V21.2td}
\end{eqnarray}

We can always choose coordinates near $\{t=0\}$ so that
\begin{equation}\label{28III22.11}
   \hat \alpha = 1+O(x)\,,
    \qquad
    \hat \beta ^A = O(x)
    \,,
\end{equation}
with $\hat \alpha$ and $\hat \beta^A$ remaining smooth.
Let $u = a + b x + c  x^{\sigma+1}/(\sigma+1)$ for some constants $a,b,c,\sigma\in \R$, $\sigma>0$.
Then
\begin{eqnarray}
  \nu &=&
    \frac{  1  }
    {  \sqrt{ 1   - ( b + c   x^{ \sigma })^2 \hat \alpha^2  }}
    \nonumber
\\ &=&
 \left\{
   \begin{array}{ll}
    \frac{  1  }
    {  \sqrt{ 1   -  b^2 \big(1+O(x)\big) }}, & \hbox{$b\ne 0\,\ \sigma \ge 1  $;} \\
    \frac{  1  }
    {  \sqrt{ 1   -  b^2 \big(1+O(x^\sigma )\big) }}, & \hbox{$b\ne 0\,\ \sigma < 1 $;} \\
    \frac{  1  }
    {  \sqrt{ 1   -  c^2  x^{2\sigma } \big(1+O(x)\big) }}, & \hbox{$b=0$;}
   \end{array}
 \right.
\\
(2)
 & = &
  O(x^2) \, \nu^3
   \,,
\\
(3)
 & = &
  O(x ) \, \nu
  \,,
\\
(1)
 & = &
   \nu\Big[ -\big(n + O(x)
   \big)
      (b+ c x^\sigma)
  \nonumber
\\
 &  &
   + x\nu^2
    \Big((\nu^{-2}+ \hat \alpha^2
   (b+ c x^\sigma)^2 ) \hat \alpha   c \sigma x^{\sigma-1}
  +\hat \alpha^2  \partial_x \hat \alpha   (b+ c x^\sigma)  ^3
   \Big)
\Big]
\\
 & = &\left\{
        \begin{array}{ll}
           \nu\Big[ -b\big(n + O(x)
    \big)
    +
 \nu^2 \big( O(x) + O(x^\sigma)\big)
   \Big)
\Big], & \hbox{$b\ne 0$;}
 \\
           \nu c x^\sigma \Big[ -\big(n + O(x)
   \big)
   +
    \Big( \sigma \hat \alpha (1+ \nu^2\hat \alpha^2
   c^2 x^{ 2\sigma }
)
  +
\nu^2 \hat \alpha^2  \partial_x \hat \alpha
  c^2  x^{2 \sigma+1 }
   \Big)
\Big], & \hbox{$b=0$.}
        \end{array}
      \right.
 \nonumber
\\
 &&
     \label{28III22.12}
\end{eqnarray}
where a more careful analysis of the error terms could be useful when  $b\ne 0$, but this is irrelevant for our further purposes.
%

When $b=0$, Equation~\eqref{28III22.12} can be rewritten as
\begin{eqnarray}
  (1)
 &=&
  cx^\sigma   \nu^3
 \Big((1 - \hat \alpha^2 c^2 x^{2\sigma})
  \big(  \sigma\hat \alpha  -   (n + O(x))\big)
+  \sigma \hat \alpha^3   c ^2  x^{2  \sigma } + O(c^2x^{2\sigma+1})
 \Big)
 \nonumber
\\   &=&
 -  cx^\sigma   \nu^3
\Big(   (1 - \hat \alpha^2 c^2 x^{2\sigma})
 \big(n+O(x)\big)
  - \sigma\hat \alpha
   + O(c^2x^{2\sigma+1})
 \Big)
 \,.
 \label{3iv22.1}
\end{eqnarray}
Collecting all this one obtains for $b=0$ and  $\hat \alpha^2 c^2 x^{2\sigma}<1$:
\begin{equation}
  H
 = - \nu^3
\Big[
  cx^\sigma\Big(   (1 - \hat \alpha^2 c^2 x^{2\sigma})
 \big(n+O(x)\big)
  - \sigma
   + O(c^2x^{2\sigma+1})
 \Big)
 + O (x )
\Big]
 \,,
 \label{3iv22.2}
\end{equation}
where the error terms are understood for small $x$.

\section{Boundary behaviour for uniformly spacelike hypersurfaces}
 \label{s3IV22.1}

Before continuing, a definition is in order:

\begin{definition}
  \label{D3IV22.1}
  A spacelike hypersurface $\hyp$ in $\mcM$ will be said to be \emph{uniformly spacelike} near $\red{\tScri}$ if its tangent space is uniformly bounded away from the null cone in a neighbourhood of  $\red{\tScri}$, in the sense that there exist Fefferman-Graham coordinates and a constant $C$ such that the graphing function $u$ of $\hyp$ satisfies
\begin{equation}\label{3IV22.21}
  \nu\equiv
    \frac{  1  + \hat \beta^i \partial_iu}
    {  \sqrt{(1  +  \hat \beta^i  \partial_i u)^2 -\hat \alpha^2 h^{ij}  \partial_i u \partial_ju }}
\le C
  \,.
\end{equation}
\end{definition}

\begin{remark}
  \label{R3IV22.2}
The bound \eqref{3IV22.21} will hold for spacelike  hypersurfaces which are differentiable up-to-boundary in the conformally rescaled spacetime, with tangent spaces which are spacelike at $\red{\tScri}$ with respect to the conformally rescaled metric.
\qed
\end{remark}

\begin{remark}
  \label{R3IV22.2+}
  Since $\nu = - \fourg( T,N)$,
it follows from the triangle inequality in the unit hyperboloid (equivalently, from the special-relativistic law of addition of velocities), that the definition is independent of the Fefferman-Graham coordinate systems when $\ovhyp\cap\tScri$ is compact.
\qed
\end{remark}

The main result of this section is:

\begin{theorem}
  \label{T3IV22.1}
Let $\hyp$ be a uniformly spacelike maximal hypersurface in \blue{$\mcM$}  such that $\ovhyp\cap \red{\red{\tScri}}$ is a smooth compact spacelike submanifold of $\red{\red{\tScri}}$. Then $\ovhyp$ is a smooth  hypersurface of $C^n$-regularity class at $\red{\red{\tScri}}$, meets $\tScri$ orthogonally, and is polyhomogeneous there.
\end{theorem}

\begin{remark}
  \label{R3IV22.1}
  The detailed differentiability properties of $\hyp$ at $\red{\red{\tScri}}$ are identical to those of the approximate solutions of Section~\ref{ss16IX21.1}.
\qed
\end{remark}

\proof
Near $\hyp\cap \red{\tScri}$ we can introduce Fefferman-Graham coordinates in which  $\hyp\cap \red{\tScri}= \{t=0\}$, with
\begin{equation}\label{28III22.11a}
   \hat \alpha = 1+O(x)\,,
    \qquad
    \hat \beta ^A = O(x)
    \,.
\end{equation}
It follows from \eqref{3IV22.22} and \eqref{3IV22.21} that there exists a constant $\slope\in [0,1)$ so that for $x$ small enough the graphing function $u$ of $\hyp$ satisfies
\begin{equation}
 |u| \le \slope x
 \,.
 \label{3IV22.23}
\end{equation}

We will need the following Lemma:
\begin{Lemma}
  \label{L26VI22.1}
  Suppose that there exists a constant $\delta <1$ such that
   the graphing function  $u$ of a maximal hypersurface satisfies \eqref{3IV22.23}
    in a Fefferman-Graham coordinate system in which
  \eqref{28III22.11a} is satisfied. Then there exist positive constants $c$, $\sigma$, and $\epsilon$ such that
\begin{equation}\label{3IV22.24}
  |u| \le \frac{c x^{1+\sigma}}{1+\sigma}
  \ \mbox{for $x\le \epsilon$}
   \,.
\end{equation}
\end{Lemma}

\begin{remark}
We note that the decay rate \eqref{3IV22.24} for \emph{some $\sigma>0$} can be improved to \emph{any $\sigma<1$}. Indeed, it follows from the already-established inequality
\eqref{3IV22.24} with some $\sigma>0$  that $\slope$ can be chosen arbitrarily small in \eqref{3IV22.23} when $\epsilon$ is chosen small enough, so that \eqref{3IV22.29} does not give any restrictions on $\sigma_0$, and hence on $\sigma$. For any $\sigma $ we can now choose $\gap$ in \eqref{4IV22.19} as close to 1 as desired. 
Hence, the error terms in \eqref{3iv22.2rt}  will be dominated by the remaining  terms   for $\epsilon$ small enough as long as  $\sigma$ is smaller than $1$.
 \qed
\end{remark} 

\noindent{\sc Proof of Lemma~\ref{L26VI22.1}:}
We claim that there exist constants $a,c,\epsilon,\sigma >0$ so that for $x\le \epsilon$ and $|a|\le \epsilon$ the graphs, say $\hyp_{a,\pm}$, of the functions $u_{a,\pm} = a \pm c x^{\sigma+1}/(\sigma+1)$ have mean curvatures
\begin{equation}\label{3IV22.19}
 \pm H[\hyp_{a,\pm}] < 0
 \,,
\end{equation}
with
\begin{equation}\label{3IV22.25}
  u_{0,-}\le u \le u_{0,+}
   \,.
\end{equation}
In order to show this, we start by increasing $\slope$ slightly if necessary,  while remaining in $[0,1)$, to obtain
\begin{equation}
 |u| < \slope x
 \
 \mbox{for $0<x\le x_0$, for some $x_0$}
 \,.
 \label{3IV22.23a}
\end{equation}
We require that $0<\sigma<\sigma_0 \red{<1}$, where $\sigma_0$ satisfies
\begin{equation}\label{3IV22.29}
  \slope  (\sigma_0+1) <1\,
\end{equation}
and that $\epsilon$ satisfies
\begin{equation}\label{3IV22.27}
 \frac{c     \epsilon^{\sigma+1}}{\sigma+1} = \slope \epsilon
  \qquad
  \Longleftrightarrow
  \qquad
 {c}   \epsilon^{\sigma } = \slope  (\sigma+1) <1
 \,.
\end{equation}
This equation defines $c$ once $\sigma$ and $\epsilon$ have been chosen.
We set
\begin{equation}\label{4IV22.19}
  \gap := 1 -
  \big(\slope (\sigma_0 +1)\big)^2 >0
  \,,
\end{equation}
and since $\sigma<\sigma_0$ we have
\begin{equation}\label{4IV22.19dr}
   1 -
    \big(
     \slope (\sigma +1)
     \big)^2
      >  \gap
  \,.
\end{equation}

It follows that there exists $\epsilon_0>0$ so that for all $|a| + \epsilon \le \epsilon_0$ we have:
\begin{enumerate}
  \item  the graphs of $u_{a,+}$ are spacelike for $0\le x\le \epsilon$, and
  \item  for $0 \le a $ the graphs of $u_{a,+}$ restricted to  $x=\epsilon$ lie  strictly above the graph of $u$, while
       \item
       for $  a\le 0$ the graphs of $u_{a,-}$ restricted to  $x=\epsilon$ lie  strictly under the graph of $u$.
\end{enumerate}

Equation~\eqref{3iv22.2} becomes
\begin{equation}
 \pm H[\hyp_{a,\pm}]
 =
  - \nu^3 \Big[
  cx^\sigma\Big(
   \underbrace{
  (1 -   \hat a ^2 c^2 x^{2\sigma} )
 \big(n+O(x)\big)
  } _{\ge \big(1 -    c^2 \epsilon^{2\sigma}  + O(\epsilon)\big)
 \big(n+O(\epsilon)\big)\ge \gap n + O(\epsilon ) }
  - \sigma
   + O(\epsilon)
 \Big)
 + O (x )
\Big]
 \,.
 \label{3iv22.2rt}
\end{equation}
 In order to  obtain  \eqref{3IV22.19}, we  reduce further  $\epsilon$ if necessary so that $\gap n +O(\epsilon)$ is larger than $\gap n/2$, and then choose  $\sigma $ smaller than $\gap n/4$, so that the coefficient of $x^\sigma$ is positive. The terms $cx^\sigma$  dominate the terms $O(x)$ for all $\epsilon$ small enough because
 $$
  \frac{cx^\sigma}{x} =c x^{\sigma-1} \ge c \epsilon^{\sigma-1}= \frac{\slope(\sigma+1)}{\epsilon}
  \,.
  $$
   We conclude that the right-hand side of \eqref{3iv22.2rt} can be made negative by further decreasing $\epsilon$, if needed.

\begin{figure}
  \centering
  \includegraphics[width=.6\textwidth]{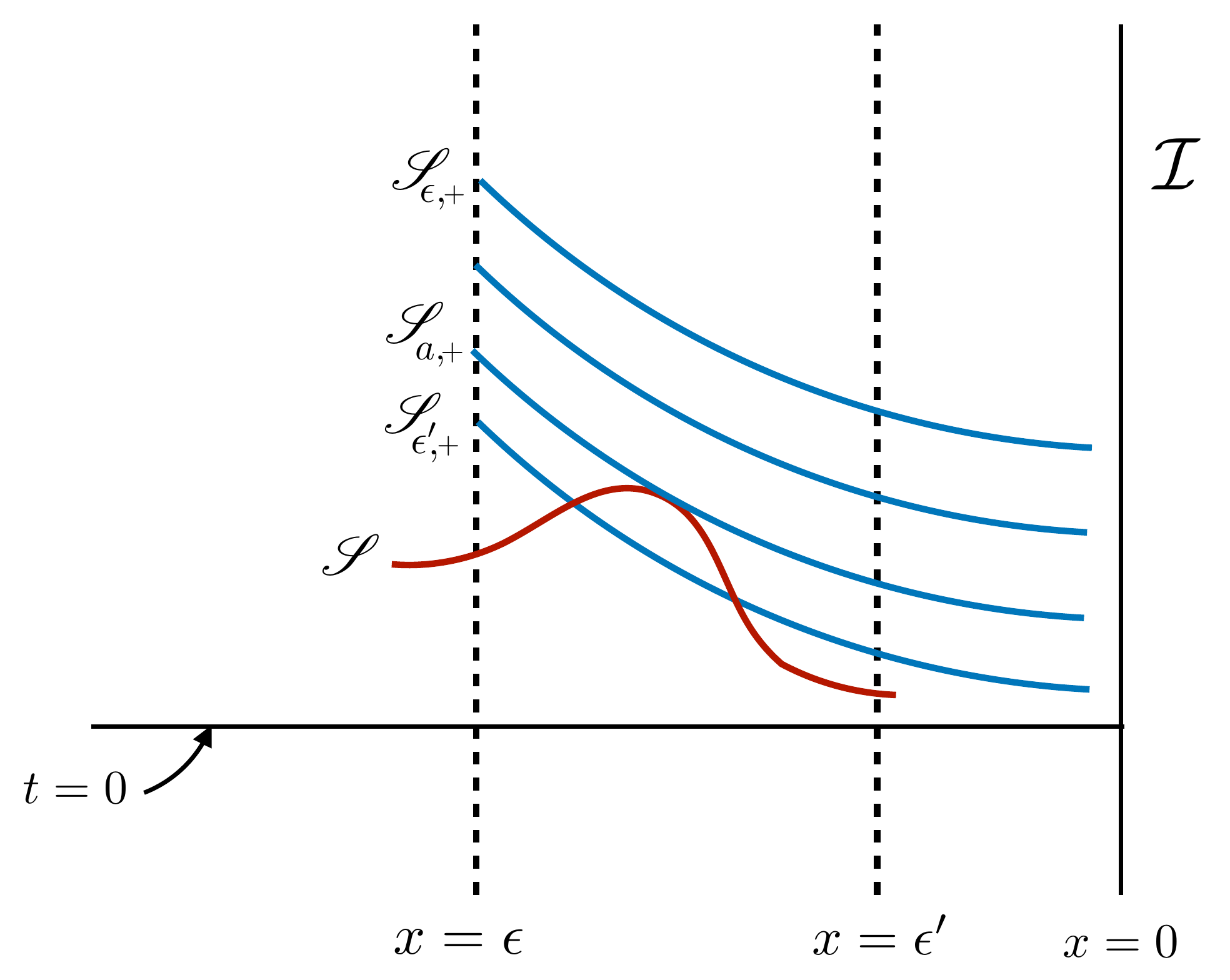}
  \caption{The barriers of the proof of Lemma~\ref{L26VI22.1}. The tangential contact  in the figure is not allowed by the maximum principle. }\label{F26VI22.1}
\end{figure}

Let $0<\epsilon'< \epsilon$ and consider the graph of $u$ over the set $\{\epsilon'\le x \le \epsilon\}$.
It holds that
$$
u_{- \epsilon,-}\big|_{\{\epsilon'\le x\le \epsilon\} }
  <
  u \big|_{\{\epsilon'\le x\le \epsilon\} }
  <
 u_{  \epsilon,+}\big|_{\{\epsilon'\le x\le \epsilon\} }
 $$
 and
  for any $a\in [ \red{\epsilon'}, \epsilon]$ we have
\begin{equation}\label{3IV22.18}
  u_{- a,-}\big|_{x=\epsilon'}
  <
  u \big|_{x=\epsilon'}
  <
  u_{ a,+}\big|_{x=\epsilon'}
  \,,
  \qquad
  u_{ - a,-}\big|_{x=\epsilon }
  <
  u \big|_{x=\epsilon }
  <
  u_{  a ,+}\big|_{x=\epsilon }
  \,.
\end{equation}
We therefore must have
\begin{equation}\label{3IV22.18a}
  u_{- \red{\epsilon'},-}\big|_{\{\epsilon'\le x\le \epsilon\} }
  <
  u \big|_{\{\epsilon'\le x\le \epsilon\} }
  <
  u_{ \red{\epsilon'},+}\big|_{\{\epsilon'\le x\le \epsilon\} }
  \,,
\end{equation}
otherwise $u$ would meet one of the surfaces of the family
$$
 \big\{
  u_{  a,+}\big|_{\{\epsilon'\le x\le \epsilon\} }
 \,,
 \
 u_{  -a,-}\big|_{\{\epsilon'\le x\le \epsilon\} }
 \big\}_{a\in [ \red{\epsilon'}, \epsilon]}
$$
tangentially, which would violate the comparison principle  for hypersurfaces with known mean curvature; see Figure~\ref{F26VI22.1}.

Passing with $\epsilon'$ to zero in \eqref{3IV22.18a} one obtains
 \eqref{3IV22.24}.
\qedskip

We return to the proof of Theorem~\ref{T3IV22.1}.
By hypothesis
the tilt function $\nu$ is uniformly bounded, hence the equation $H=0$ is uniformly elliptic.   Interior elliptic estimates for equations in divergence form together with \eqref{3IV22.24} give $\partial u = O(x^{\sigma})$, and
the result follows from Theorem~\ref{T19IX21.1}.
\qed

\section{Uniqueness}
 \label{s10V21.1}

Let $(\mcM,g)$ be locally asymptotically hyperbolic
and let the closure $\ovhyp$ of $\hyp$ be a compact Cauchy surface in the conformally completed spacetime.
 Near the conformal boundary $\tScri$ we use coordinates such that $\hyp = \{t=0\}$.

 Consider two smooth spacelike Cauchy surfaces $\hyp_1$ and $\hyp_2$ in the globally hyperbolic region $\mcD(\hyp)$, such that that $\partial\ovhyp\!_1 = \partial\ovhyp\!_2 = \partial\ovhyp$.


 For  $q\in \hyp_2$ let $f_{\hyp_1}$ be defined as
   \begin{equation}\label{11IX21.31}
     f_{\hyp_1}(q):= d(\hyp_1,q) \equiv \sup_{p\in \hyp_2}d(p,q)
     \,,
   \end{equation}
   where $d(p,q)$ is the Lorentzian distance from $p$ to $q$ (zero if $q\not \in J^+(p)$). Under the present assumptions,
$f_{\hyp_1}$ is continuous.

 \begin{Lemma}
   \label{L11IX21.1}
   Let $u_a$, $a=1,2$ be the graphing functions for $\hyp_1$ and $\hyp_2$.   If  $u_a = o(x)$,  $a = 1,2$,
 then for every $\epsilon>0$ there exists $\delta>0$ such that if $x(q)<\delta$ then $f_{\hyp_1}(q)<\epsilon$.
 \end{Lemma}

\proof
Note that $f_{\hyp_1}(q) = 0$ if $q\in \hyp_2\cap J^-(\hyp_1)$, in which case there is nothing to prove. It remains to consider $q\in \hyp_2\cap J^+(\hyp_1)$.

Let us start by showing that  for points $R=(r,u(r))$,
 with $r\in \hyp$ sufficiently close to the boundary,
  and  with $0<u(r) = o(x(R))$ we have
$$
 d(\hyp, R)  \le C u(r)/x (R)\,.
$$
For this, let $s\mapsto\gamma(s) $ be any timelike curve from $\hyp$ to $R$ with tangent denoted by $\dot \gamma$, we have
\begin{eqnarray}
 L(\gamma)
  & = &
 \int x^{-1}\sqrt{ \hat \alpha^2 \dot t^2 - \dot x^2
 - h_{AB}(\dot x^A + \hat \beta ^A\dot t) (\dot x^B + \hat \beta ^B\dot t)
 } ds
 \nonumber
\\
 & \le &
 \int x^{-1}  \hat \alpha  \dot t ds
  \le C x(R)^{-1} u(r)
  \,,
 \label{11IX21.33}
\end{eqnarray}
since $x(\gamma(s))\ge x(R)/C_1$ for some constant $C_1$ for all future directed timelike curves from $\hyp$ to $R$ when sufficiently close to the boundary. Hence
\begin{eqnarray}
  d(\hyp, R) = \sup_\gamma L(\gamma)
 & \le &  C x(R)^{-1} u(r)
  \,.
 \label{11IX21.34}
\end{eqnarray}
Similarly one shows that for $u(r)<0$
\begin{eqnarray}
  d(R,\hyp) = \sup_\gamma L(\gamma)
 & \le &  C x(R)^{-1} |u(r)|
  \,,
 \label{11IX21.34a}
\end{eqnarray}
where now the sup is taken over  timelike curves from $R$ to $\hyp$.

Let $R=(r,u_2(r))\in \hyp_2$, and let $Q=(q,u_1(q))$ maximise the distance between $\hyp_1$ and  $R$; thus $Q\in  J^-(R)$.

\begin{figure}
  \centering
  \includegraphics[width=.9\textwidth]{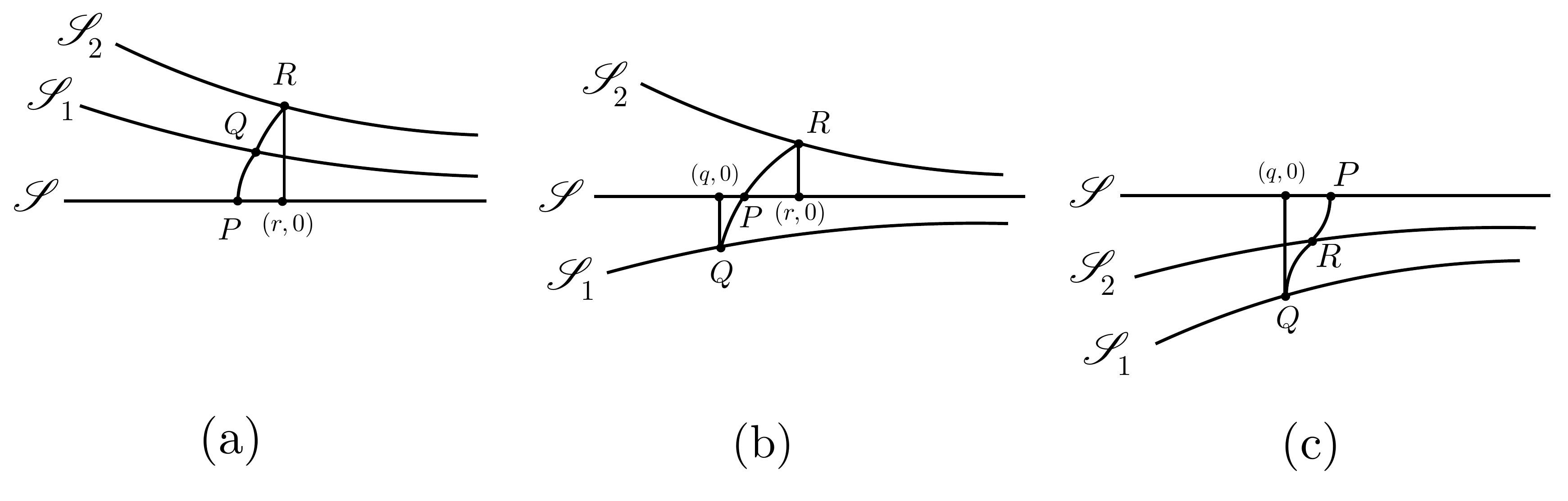}
  \caption{ (a) First case; (b) second case; (c) third case.}\label{F11IX21.1}
\end{figure}

Assume, first, that $u_2(r)>0$ and $u_1(q)\ge 0$; see the case (a) of Figure~\ref{F11IX21.1}.
%
%
For any point $P \in \hyp \cap J^-(Q)$,
\begin{eqnarray}
 d(\hyp_1,R) &=&   d(Q,R) \le d(P,Q)+d(Q,R)  \nonumber \\
&\le&  d(\hyp,R)
\le C  x(R)^{-1}  u_2(r) \, ,
\end{eqnarray}
where in the last step we have used \eqref{11IX21.33} with $u$ replaced by $u_2$.


Assume, next, that $u_2(r)\ge 0$ and $u_1(q)\le 0$, see Figure~\ref{F11IX21.1} (b).
%
Then there exists a point $P \in \hyp$ such that,
\begin{eqnarray}
   d(\hyp_1,R)&=&d(Q,P) + d(P,R) \le d(Q,\hyp) + d(\hyp,R)
    \nonumber
\\
 &
    \le
    &
     C
    \big(
     x(Q)^{-1} |u_1(q)| +
     x(R)^{-1} u_2(r)
     \big)
     \,.
     \label{11IX21.51a}
\end{eqnarray}

It remains to analyse the case  $u_2(r) <0$ and $u_1(q)< 0$, see case (c) of Figure~\ref{F11IX21.1}.
For any point $P \in \hyp \cap I^+(R)$,
\begin{eqnarray}
   d(\hyp_1,R)  &=& d(Q,R)  \le   d(Q,R) + d(R,P) \nonumber \\
  &\le&  d(Q,\hyp)   \le   C x(Q)^{-1} |u_1(q)|     \,.
     \label{11IX21.51b}
\end{eqnarray}
The result follows.
\qed

\bigskip

Our main result in this section is the following:

\begin{theorem}
 \label{T7IV22.1}
Let the setting be as in the
first paragraph of the current section
where, in addition, we require that
$(\mcM,g)$ satisfies the timelike convergence condition:
$$
{\rm Ric}(X,X) \ge 0 \quad\mbox{for all timelike vectors } X\,.
$$
Let $u_1$ and $u_2$  be the graphing functions for  $\hyp_1$ and $\hyp_2$ with $u_1\big|_{x=0}=u_2\big|_{x=0}$.
Suppose that  for small $x$, we have
\begin{equation}
 \label{23VII22.1}
u_a= o(x)\,, \quad a=1,2
\,.
\end{equation}
If $\hyp_1$ and $\hyp_2$ are maximal (i.e.\  have mean curvature zero) then $\hyp_1 = \hyp_2$.
\end{theorem}

\begin{remark}
  \label{R15VII22.1}
  The result remains true, with the same proof, if $\hyp_1$ and $\hyp_2$ share a common boundary within $\mcM$.
\end{remark}

\begin{remark}
When $(\mcM,g)$ satisfies the Einstein equations with cosmological constant $\Lambda$, the spacetime Ricci tensor is given by
$$
R_{\mu\nu} = T_{\mu\nu} -  \frac1{n-1} T g_{\mu\nu} + \frac{2\Lambda}{n-1} g_{\mu\nu}  \,.
$$
In particular, a {\it negative} cosmological constant helps to promote the timelike convergence condition,
since $g$ is negative on timelike vectors in our signature.
\qed
\end{remark}

\begin{cor}
  \label{C23VII22.1}
Under the  hypotheses of Theorem~\ref{T7IV22.1} other than \eqref{23VII22.1}, suppose that
\begin{enumerate}
 \item either   the $\hyp_a$'s are both  differentiable and spacelike \emph{up-to-boundary} in the conformally rescaled metric on $\bmcM$,
     \item
     or the manifolds $\hyp_a$ are uniformly spacelike in the sense of Definition~\ref{D3IV22.1} (compare Remark~\ref{R3IV22.2}),
\item
or there exists $\slope \in (0,1)$ such that $|u_a|\le \slope x$ in Fefferman-Graham coordinates in which $\hat \alpha|_{x=0}=0=\hat \beta^A|_{x=0}$.
\end{enumerate}
If $\hyp_1$ and $\hyp_2$ are maximal,  then $\hyp_1 = \hyp_2$.
\end{cor}

\noindent{\sc Proof of Corollary~\ref{C23VII22.1}:}

 1. Proposition~\ref{P4IV22.1} shows that \eqref{23VII22.1} is satisfied.

2. Theorem~\ref{T3IV22.1} shows  that  \eqref{23VII22.1} is satisfied.

3. Lemma~\ref{L26VI22.1} shows that  \eqref{23VII22.1} is satisfied.

In each case we can thus apply Theorem~\ref{T7IV22.1}.
\qedskip

\noindent {\sc Proof of Theorem~\ref{T7IV22.1}:}
Suppose $\hyp_1 \ne \hyp_2$. Then, without loss of generality, we may assume that $\hyp_2$ meets
$I^+(\hyp_1)$. With the aid of  Lemma \ref{L11IX21.1}, one can then establish the existence of points $p \in \hyp_1$ and $ q \in \hyp_2$ such that $d(p,q) = d(\hyp_1, \hyp_2)$.  Let  $\gamma: [0,\ell] \to \mcM$ be a unit speed timelike geodesic from $\g(0) = p$ to $\g(\ell) = q$  having length $d(p,q)$ ; $\g$ necessarily meets $\hyp_1$ and $\hyp_2$ orthogonally.  Then by \cite[Theorem 3]{Eschenburg}, a neighbourhood $V$ of $\g$ splits.  More precisely, there exists a neighbourhood $U$ of $p$ in $\hyp_1$ such that
$\psi: [0,\ell] \times U \to V$ defined by $\psi(s,x) = \exp_x (sT)$, where $T$ the future direct unit normal to $\hyp_1$, is an isometry, with $\psi(\ell,U)$ a neighbourhood of $q$ in $\hyp_2$.
(Here, $[0,\ell] \times U$ carries the product metric $-dt^2 \oplus h|_U$, where $h$ is the induced metric on $\hyp_1$.)
By a straightforward continuation argument $\psi$ extends to an isometry $\psi:  [0,\ell] \times \hyp_1 \to J^+(\hyp_1) \cap J^-(\hyp_2)$,  $\psi(s,x) = \exp_x sT$,
where $[0,\ell] \times \hyp_1$ carries the product metric, i.e.\  the region between $\hyp_1$ and $\hyp_2$ splits  isometrically as a product.
We can apply   Lemma \ref{L11IX21.1}  and a contradiction results.
\qed

\section{Local uniqueness and regularity}
 
The aim of this section is to prove that maximal hypersurfaces that satisfy the slope bound of Lemma~\ref{L26VI22.1}
meet the boundary orthogonally and are regular there when the timelike convergence condition holds:

\begin{theorem}
  \label{T3IV22.2}
Let $\hyp$ be a maximal hypersurface in an ALH spacetime $(\mcM,\fourg)$ such that $\ovhyp\cap \red{\red{\tScri}}$ is a smooth compact spacelike submanifold of $\red{\red{\tScri}}$.
Assume, for some $\slope \in [0,1)$,
$\hyp$ satisfies \eqref{3IV22.23} in a Fefferman-Graham coordinate system in which \eqref{28III22.11a} holds.
If  $(\mcM,\fourg)$ obeys the timelike convergence condition, then there exists a constant $C$ such that the graphing function $u$ of $\hyp$ is polyhomogeneous and satisfies,
 for small $x$,
\begin{equation}\label{3IV22.24-}
  |u| \le C x^{2}
   \,.
\end{equation}
\end{theorem}

\begin{Remark}
  \label{R26VI22.1}
The timelike convergence condition is used to guarantee uniqueness of solutions of the local Dirichlet problem for maximal hypersurfaces near the conformal boundary.  It is conceivable that local uniqueness near enough to the boundary holds without further conditions in our setting, in which case the timelike convergence condition would not be needed; in any case, the result remains true with any other condition which guarantees uniqueness for such solutions.
\qed
\end{Remark}

{\sc Proof of Theorem~\ref{T3IV22.2}:}
Let us denote by  $\epsilon_0>0$ a constant such that   the hypersurfaces $\hyp_{a,\pm}$ of the proof of Lemma~\ref{L26VI22.1} are barriers for $x\le \epsilon_0$.
By that last lemma there exists $\sigma>0$ so that $u= O(x^{1+\sigma})$ for small $x$.

Decreasing $\epsilon_0$ if necessary we can assume that the domain of dependence $\mcD(\hyp\cap \{0\le x\le \epsilon_0\}, \ovmcM)$ has compact closure.

There exists $\epsilon < \epsilon_0$ so the graph of $u$ restricted to $x=\epsilon$ is contained in the interior of $\mcD(\hyp\cap \{0\le x\le \epsilon_0\}, \ovmcM)$.

Consider a sequence of maximal hypersurfaces, say $\hyp_i$, which are graphs of  solutions $u_i$  of the maximal hypersurface equation with boundary values $u_i|_{x=\epsilon}=u|_{x=\epsilon}$ and $u_i|_{x=1/i}=0$. Existence of the $u_i$'s follows from e.g. \cite[Theorem~4.2]{Bartnik84}, where the condition of compactness of the relevant domain of dependence is enforced by the last choice of $\epsilon$.

It follows from~\cite{Seifert}%
\footnote{For completeness we give a simple construction of $\tilde t$ in Appendix~\ref{app13VII22.1}.}
that there exists a time function $\tilde t$, defined  in a compact neighborhood of all the graphs $u_i$, so that a)  the graphing function of $\hyp_i$, say $\tilde u_i$, with respect to this time function has zero boundary values both at $x=\epsilon$ and $x=1/i$, and b) such that $t$ coincides with $\tilde t$ for $x<\eta$ for some $\eta>0$.
The barriers of Lemma~\ref{L26VI22.1} show that the $\tilde u_i$'s decay faster than $Cx^{1+\sigma}$, for some constant $C$ independent of $i$, hence satisfy the hypotheses of Theorem~\ref{T19III22.1}, and therefore have a tilt function bounded independently of $i$.

Passing  to the limit we obtain a maximal hypersurface whose   graphing function $\tilde u_\infty$
 has bounded tilt and is $O(x^{1+\sigma})$.

The graph of $\tilde u_\infty$ coincides with the graph of $u$ at $x=\epsilon$, and both graphs approach $t=0$ faster than $x$, so that  point 1. of Theorem~\ref{T7IV22.1} applies (compare Remark~\ref{R15VII22.1}).  Hence the graph of $\tilde u_\infty$ coincides with the graph of $u$. Since
$\tilde u_\infty$ has bounded tilt, so does $u$. We can thus apply Theorem~\ref{T3IV22.1} to reach the desired conclusion.
\qedskip

We shall say that \emph{local uniqueness} holds if
    there exists $\epsilon>0$ so that solutions over $\{0<x<\epsilon\}$ of the Dirichlet problem for the prescribed mean curvature equation with smooth data at $x=\epsilon$ and which decay at least as $x^{1+\sigma }$ for some $\sigma>0$ are unique.
The reader might have noticed that the above arguments establish the following version of Akutagawa estimates, which does not assume compactness of $\Omega$:

\begin{Proposition}
  \label{P26VI22.3}
  Let  $u$ be the graphing function of a maximal hypersurface over
   a  domain $\Omega \subset \{t=0\}$ with smooth boundary and with compact closure in $\omcM$, where $t$ is, near the conformal boundary, part of a Fefferman-Graham coordinate system in a conformally completed ALH spacetime.
   Assume that $\partial \Omega$ has a component coinciding with $\{t=0\}\cap \scri$.
   Let the constant $C>0$ be such that  along the graph we have
\begin{eqnarray}
 &
   \displaystyle
   |\Ric |_{\backriemg} +  |  \mcL_T \fourg|_{\backriemg}
   +  |\nabla \mcL_T \fourg|_{\backriemg}
   +  |\alpha^{-1}   \nabla \alpha|_{\backriemg}
   +  |\alpha^{-1} \nabla \nabla \alpha|_{\backriemg}
   \le C
   \,.
   &
    \label{26VI22.41}%
\end{eqnarray}
If moreover $\partial \Omega\cap \mcM\ne \emptyset$,   let $C_1$ be such that
\begin{equation}\label{27III22.1a}
  | H_{\partial \Omega\cap \mcM}|_{\backriemg} \le C_1
  \,.
\end{equation}
If local uniqueness holds, and if there exist positive constants $\delta$ and $\hat C$  such that in Fefferman-Graham coordinates we have
\begin{eqnarray}
 &
   \displaystyle
   |u | \le \hat C x ^{1+\delta}
   \,,
    \label{20III22.2a}
   &
\end{eqnarray}
then there exists a constant $\tilde C(C,C_1,\hat C, \delta)$ such that
\begin{eqnarray}
 &
   \displaystyle
   \nu \le \tilde C
   \,.
    \label{27III22.2a}
   &
\end{eqnarray}
\qed
\end{Proposition}

\section{Existence results}
 \label{11IX21.1}

The results about existence of maximal hypersurfaces in ALH spacetimes are scarce in the literature.
Akutagawa~\cite{Akutagawa} proved existence of a maximal hypersurface in three dimensional AH spacetimes under the hypothesis of existence of uniformly spacelike barriers.
The generalisation of his result  to all topologies and dimensions is essentially trivial, and is made explicit in Theorem~\ref{T11IX21.1a} below.
In~\cite{ShiMax}  the implicit function theorem was used to prove existence of maximal hypersurfaces for metrics near the Anti-de Sitter one in all dimensions. We point out below that this generalises to all conformally compactifiable asymptotically vacuum static ALH
metrics satisfying an energy condition, in all dimensions, cf.\ Theorem~\ref{T4X21.1}.

Recall that a standard method for constructing solutions of elliptic nonlinear PDEs is that of barrier functions. Typically a barrier will be a function for which the equality in the equation is replaced by an inequality. In our context we will need to impose some further conditions on the barriers, which will allow us to control the solutions near the boundary. Thus, we will say that an acausal spacelike hypersurface $\hyp$ is a \emph{good barrier} when its conformal completion  is compact and if one of the following conditions hold (compare Theorem~\ref{T7IV22.1}):

\begin{enumerate}
 \item there exists a Fefferman-Graham coordinate system in which $\hat\alpha|_{x=0}=1$ and $\hat \beta^A|_{x=0}=1$ so that, for small $x$, the graphing function   $u $ of $\hyp$ satisfies $ u = \slope x + o(x)$ for some $\slope \in (-1,1)$,
     \item
     or   $\hyp $ is uniformly spacelike in the sense of Definition~\ref{D3IV22.1} (compare Remark~\ref{R3IV22.2}),
 \item or  $\hyp $ is  differentiable and spacelike \emph{up-to-boundary} in the conformally rescaled metric on $\bmcM$,
\end{enumerate}

As such, the key condition for our purposes is the first one.
We note that the third condition implies the second, and the second implies the first.

\begin{theorem}
  \label{T11IX21.1a}
 Let $(\mcM,\fourg)$ be  an ALH spacetime such that $(\bmcM,\widetilde \fourg)$ is globally hyperbolic (in the sense of manifolds with timelike boundary) with compact Cauchy surfaces, and let $H_1\in C^\infty(\omcM)$, with $H_1|_{x=0}=0$.
Suppose that there exist good barriers $\hyp^\pm$, sharing a common boundary at conformal infinity,
with mean curvatures $H^\pm$ satisfying
$$
H^- \ge H_1 \ge H^+
 \,,
$$
and with
$$
 \ovhyp ^+\subset J^+(\ovhyp^-,\omcM)
 \,.
$$
Then there exists    a spacelike hypersurface $\hypone \subset    \mcM$, with mean curvature equal to $H_1$, such that
$$
 \ovhypone \subset  J^+(\ovhyp^-,\omcM) \cap   J^-(\ovhyp^+,\omcM)
  \,.
$$
\end{theorem}

%

\smallskip

\proof
The proof is a repetition of that in~\cite{Akutagawa}.
Lemma~\ref{L26VI22.1}
provides \eqref{3IV22.24}.
 The a priori estimate on the tilt function $\nu$ needed for the proof is provided by   Theorem~\ref{T19III22.1}; we check that the hypotheses of Theorem~\ref{T19III22.1} are satisfied in the current setting in Appendix~\ref{App6IV22.1}.
\qedskip

Recall that a metric $\fourg$ is said to satisfy the \emph{timelike convergence condition} if for all causal vectors $X$ we have
\begin{equation}\label{10IX22.1}
  {\bf\Ric}(X,X) \ge 0
  \,,
\end{equation}
where ${\bf \Ric}$ is the Ricci tensor of $\fourg$.
 We have:

\begin{theorem}
  \label{T28VII22.1}
  Under the hypotheses of Theorem~\ref{T11IX21.1a}, suppose in addition that $H_1 \equiv 0$ and that the timelike convergence condition is satisfied by the metric $\fourg$.
  Then the hypersurface constructed in that theorem is uniquely defined by its boundary values, is polyhomogeneous, smooth in the interior, and is of differentiability class $C^n$-up-to-boundary at $\tScri$,
  where $n+1$ is the dimension of spacetime.
\end{theorem}

\proof
The hypotheses of Theorem~\ref{T11IX21.1a} guarantee that Theorem~\ref{T3IV22.2} applies, and regularity follows from that last theorem; compare Remark~\ref{R3IV22.1}. Uniqueness follows from Theorem~\ref{T7IV22.1}.
\qedskip

The following result encompasses the existence theorem of~\cite{ShiMax}:

\begin{Theorem}
 \label{T4X21.1}
Consider an ALH manifold $(\mcM,\mathring \fourg)$ satisfying the timelike  convergence
 condition such that the conformally rescaled metric on $\omcM$ is  smooth up-to-boundary and globally hyperbolic.
Suppose that there exists a coordinate system on $\omcM$, as in \eqref{9V21.6} near $\tScri$, with the level sets of $t$ maximal and compact. Then for every metric such that $x^{2} \fourg$ is smooth and sufficiently close to $x^2 \mathring \fourg$ in $C^7(\ovhyp)$, and for every smooth boundary function $\psi$ sufficiently close to zero in $C^3(\ovhyp)$,
there exists a maximal hypersurface $\hyp_0$ such that
$$
 \ovhyp_0 \cap \partial \mcM  = \{ t= \psi\}
 \,.
$$
The graphing function $u$ of $\ovhyp_0$ satisfies $u-\psi=O(x  )$ for small $x$, and is polyhomogeneous.
\end{Theorem}

\begin{remark}
  The level sets of $t$ will be maximal when $\mathring \fourg$ is, e.g., static.
  \qed
\end{remark}

\proof
We use the notation for functional spaces of \cite{AndChDiss}.

On $\hyp$ we use coordinates $(x, x^A)$ induced from a Fefferman-Graham coordinate system $(x,t,x^A)$ for $\mathring\fourg$ such that $\hyp = \{t=0\}$ near $\tScri$.
Writing
\begin{equation}\label{27VII22.1}
  \mathring \fourg = - \mathring\alpha^2 dt^2  + \mathring g_{ij}(dx^i +\mathring\beta^i dt)(dx^j+\mathring \beta^j dt)
  \,,
\end{equation}
we further require, for simplicity, that  $\lim_{x\to 0}x^2\mathring \alpha =1$ and $\lim_{x\to 0} \mathring \beta ^A =1$.

Let $\psi\in C^{k+\sigma} (\ovhyp\cap\tScri)$, $k\ge 2$, $\sigma \in (0,1)$, and let $\hat\psi \in C^{k+\sigma}(\ovhyp)$ be obtained by extending $\psi$ from $\ovhyp\cap\tScri$ to $\ovhyp$ to a function on $\hyp$ with   $x$-derivatives up to order $k$ vanishing at $\ovhyp\cap\tScri$.

Let $\bchi \in  x^{-2}C^{k-1+\sigma}(\ovhyp)$ be a symmetric tensor field with $\bchi_{x x^\mu}=0$ near the boundary, with  $\lim_{x\to 0}x^2\mathring \bchi_{tt} =1$ and $\lim_{x\to 0} x^2 \mathring \bchi_{tA} =1$, and with small norm so that the tensor field $\fourg=\mathring \fourg + x^{-2} \bchi$ has Lorentzian signature. Note that the condition $\bchi_{x x^\mu}=0$ puts $\fourg$ in the Fefferman-Graham form, and that we have $\hat \alpha|_{x=0}=1$, $\hat \beta^A|_{x=0}=0$. We emphasise that every ALH Lorentzian metric near $\mathring\fourg$ can be put in this form by a  choice of coordinates, with  a loss of not more than six derivatives for the metric functions.%
\footnote{Some comments might be in order here. There is a loss of no more than two orders of differentiability in $C^{k+\sigma}$ spaces when introducing   Gauss coordinates associated with $\ovhyp\cap\tScri$ at the conformal boundary. It follows from~\cite[Lemmata~5.1 and A.1]{Lee:spectrum} that there is a further loss of no more than two orders of differentiability for the Fefferman-Graham coordinate functions, hence no more than three orders of differentiability for the metric coefficients. Altogether an ALH metric which is in $C^{\ell +6}\supset C^{5+\ell+\sigma}$ after conformal rescaling will have metric coefficients which are at least of $C^{ \ell+\sigma}$ differentiability class in Fefferman-Graham coordinates. This estimation of the loss of derivatives can be substantially reduced for polyhomogeneous metrics (cf., e.g., \cite[Lemma~6.1]{CDLS}) if needed, but this is of no concern to us here.}
For $\hat u\in  C^{0 }_{k+\sigma}(\hyp) $ consider the map
\begin{equation}\label{26VII22.1}
  (\psi,\chi,\hat u)\mapsto  H[\hat\psi+ x\hat u]|_{\fourg= \mathring \fourg + x^{-2} \bchi}
   \in  C^{0}_{k-2+\sigma}(\hyp)
  \,,
\end{equation}
where the norms of  the fields $(\psi,\chi,\hat u)$ are small in their respective spaces.
When \eqref{10IX22.1} holds  the linearisation of \eqref{26VII22.1} with respect to $\hat u$ at $\psi=\bchi=0$ is an isomorphism  in weighted Sobolev spaces with the indicated decay rates, see the references in the proof of Theorem~\ref{T19IX21.1}.
The result follows from the implicit function theorem.
%
\qed

\medskip
%

%
%

\bigskip

We finish this work with an existence result similar in spirit to that of~\cite{BonsanteSchlenkerTeichmuller}. In that last work maximal hypersurfaces in AdS spacetimes with Dirichlet data at conformal infinity are constructed. The argument gives existence, but both here and in~\cite{BonsanteSchlenkerTeichmuller}  no information about differentiability at the boundary at infinity is provided. (Strictly speaking, the hypersurfaces of \cite{BonsanteSchlenkerTeichmuller} have some more regularity than claimed here, namely bounded extrinsic curvature, but whether or not this suffices to control differentiability at $\tScri$ remains to be seen.)

\begin{theorem}
  \label{T11IX21.1}
Let $(\mcM,\fourg)$ be an  ALH spacetime with a globally hyperbolic conformal completion  $(\omcM,\tilde \fourg)$,  and let  $\hyp$ be  a partial Cauchy surface in $\mcM$.   Suppose  that $\hyp$    has compact closure $\ovhyp$   in $\omcM$ and intersects the conformal boundary of $\mcM$ in a smooth  spacelike submanifold $\partial \ovhyp$.
If
$$
\mbox{the closure in $\omcM$ of the domain of dependence $\mcD(\hyp,\mcM)$   of $\hyp$ is compact,}
$$
%
%
then for every smooth function $H_1\in C^\infty(\mcM)$ there exists a spacelike hypersurface $\hypone \subset  \mcD(\hyp,\mcM)$,
Cauchy for $\mcD(\hyp,\mcM)$,
smooth in $ \mcM $, with mean curvature equal to $H_1$, such that
$$
 \ovhypone \cap \red{\tScri} =
 \ovhyp \cap \red{\tScri}
  \,.
$$
\end{theorem}

 \begin{Remark}
  \label{R4XI22.1}
There are natural generalisations of all the results in this section, where the prescribed-mean-curvature hypersurface has interior boundaries.
The  hypersurfaces constructed will be as smooth in $\mcM$ as the metric allows, e.g.\ smooth if the metric is smooth.

For instance,
 Theorem  \ref{T11IX21.1} remains valid if the assumption that  $\hyp$ is  a partial Cauchy surface in
 $\mcM$ is replaced by the assumption that $\hyp$ is an acausal spacelike hypersurface in  $\mcM$ such that  $\ovhyp \cap \mcM$ has smooth boundary with spacelike tangent spaces.  This compact boundary (perhaps with multiple components) will be shared by the hypersurface that we construct.
 %
\qed
\end{Remark}

\begin{Remark}
  \label{R15IV22}
 We are allowing   functions $H_1$ which are smooth on $\mcM$, without any restrictions on the behaviour of $H_1$ as   $\tScri$ is approached.  Clearly regularity at $\tScri$ would require some such conditions on $H_1$. But even in the case $H_1\equiv 0$ the argument below does not provide direct information about the behaviour of the resulting maximal hypersurface at $\tScri$,
as needed e.g.\ to define the renormalised volume
 \qed
\end{Remark}

%
%

\begin{remark}
  \label{R11IX21.1}
An identical existence result holds for metrics such that the conformally rescaled metric is differentiable on $\omcM$ and polyhomogeneous.
\qed
\end{remark}

\begin{remark}
 In the maximal case, if one assumes that the timelike convergence condition holds, an argument similar to that used in proving uniqueness shows that the maximal hypersurface so constructed is of maximal volume with respect to compactly supported variations.
 \qed
\end{remark}

\proof
In the coordinates of \eqref{9V21.6}, chosen so that $t=0$ on $\hyp$ for small $x$,
for $i\in \N$ we  let
$$
 \hyp_i=\hyp\setminus \{x< 1/i\}
 \,.
$$
Then the   closure of the
domain of dependence of $\hyp_i$ is
a  compact
subset of  $\mcD(\hyp,\mcM)$.
By~\cite[Theorem~4.1]{bartnik:variational}
there exists a smooth, acausal and spacelike hypersurface $\whyp_i$ with boundary coinciding with $\partial\hyp_i$,  with mean curvature $H$ equal to $H_1$,
and which is contained in the domain of dependence of $\hyp_i$.

%
%

Let us denote by $t$  a   time function on {$\omcM$} as constructed in Appendix~\ref{app13VII22.1}, so that $t$ vanishes on $\hyp$ and coincides with a Fefferman-Graham time coordinate near $x=0$ and $t=0$.
Let $u_i$ denote the graphing function of $\whyp_i$ over $\hyp_i$, with respect to $t$. We extend $u_i$ to a function defined over $\hyp$ by setting $u_i\equiv 0$ for $x\le 1/i$.

 The following result is standard, we give a detailed presentation for completeness; we note that global hyperbolicity of $(\omcM,\tilde \fourg)$ is not needed for its proof as presented here:

\begin{Lemma}
  \label{L27VII22.1} There exists a subsequence  $\{u_{i_k}\}_{k\in \N}$  which converges uniformly over any compact set to an achronal graph over $\hyp$.
\end{Lemma}
\proof

In order to avoid the need of considering separately the  boundary points of   $\whyp_i$ in the arguments below, we first extend $\omcM$ and $x^2\fourg$ across $\tScri$ near $\ovhyp\cap\tScri$ in any smooth way. We extend  $\hyp$ continuously across $\tScri$ to a closed  acausal hypersurface, which we denote by $\thyp$. Next we continuously extend $\whyp_i$ across its boundary by replacing  within $\thyp$ the hypersurface $\hyp_i$ by $\whyp_i$, and denote by $\wtthyp_i$ the extended surface.

 Now, let
  $p\in \overline{\mcD(\hyp ,\mcM)}$,
  and let  $\mcU_p$ be a  coordinate patch near $p$ with coordinates $(\tau, y^i)\equiv (\tau,\vec y)$, where the time coordinate $\tau$ runs  over $(-8 r_p,8 r_p)$ and the space coordinates $\vec y $ range  over a coordinate ball $B_p(4 r_p)$ of radius $4 r_p$. The coordinates on $\mcU_p$ are chosen so that the coordinate-slopes of the light cones are bounded from above by $2$ and below by $1/2$. We denote by $\mcV_p$ the subset of $\mcU_p$ coordinatised by  $(\tau, y^i)\in (-r_p,r_p)\times B_p(r_p )$. Then the $\tau$-projection of every acausal spacelike hypersurface which is closed in $\mcU_p$ and which meets $\mcV_p$ covers $B_p(2r_p )$.

Consider 
the $\tau$-graphing function of $\wtthyp_i\cap \mcV_p$, if non-empty.
  The intersection $\wtthyp_i\cap \mcV_p$ is acausal, hence a uniformly Lipschitz graph over $B_p(2r_p)$,
   with a coordinate Lipschitz-bound $2$.
    It follows that for every $p$ such that $\mcV_p$ contains an accumulation point of the $\wtthyp_i$'s, there exists a subsequence which converges to a Lipschitz hypersurface within $\mcV_p$.

By compactness a finite number $\mcV_{p_a}$, $a=1,\ldots,N$, of $\mcV_p$'s covers $\overline{\mcD(\hyp,\mcM)}$. From this collection we discard those which do not contain accumulation points of the $\wtthyp_i$'s, and we reorder them so that the first $N_1$ contain such points.

From what has been said, there exists a  sequence $\{i_j\}_{j\in\N}$  such that $\wtthyp_{i_j}\cap \mcV_{p_1}$ converges to a Lipschitz graph $\wtthyp_{1,\infty}$ within $\mcV_{p_1}$. We  set $\wtthyp_{1,j}=\wtthyp_{i_j}$.

Next, there exists a  sequence $\{j_k\}_{k\in\N}$  such that $\wtthyp_{1,j_k}\cap \mcV_{p_2}$ converges to a Lipschitz graph $\wtthyp_{2,\infty}\subset\mcV_{p_2}$. We  set $\wtthyp_{2,k}=\wtthyp_{1,j_k}$. Note that $\wtthyp_{2,\infty}\cap \mcV_{p_1}$ coincides with $\wtthyp_{1,\infty}\cap \mcV_{p_2}$.

Continuing  in this way, after $N_1$ steps we obtain a subsequence $
\{\wtthyp_{N_1,j}\}_{j\in \N}$ of $\{\wtthyp_{i}\}_{i\in\N}$ such that  $\wtthyp_{N_1,j}\cap \mcV_{p_a}$ converges, as $j$ tends to infinity, to the same hypersurface   $\wtthyp_{a,\infty}$ within $\mcV_{p_a}$ for each $a\in \{1,\ldots,N_1\}$.
The result is obtained by graphing the hypersurface $\cup_{a=1}^{N_1} \wtthyp_{a,\infty}$.
\qedskip

\begin{figure}
\begin{center}
\includegraphics[scale=0.33]{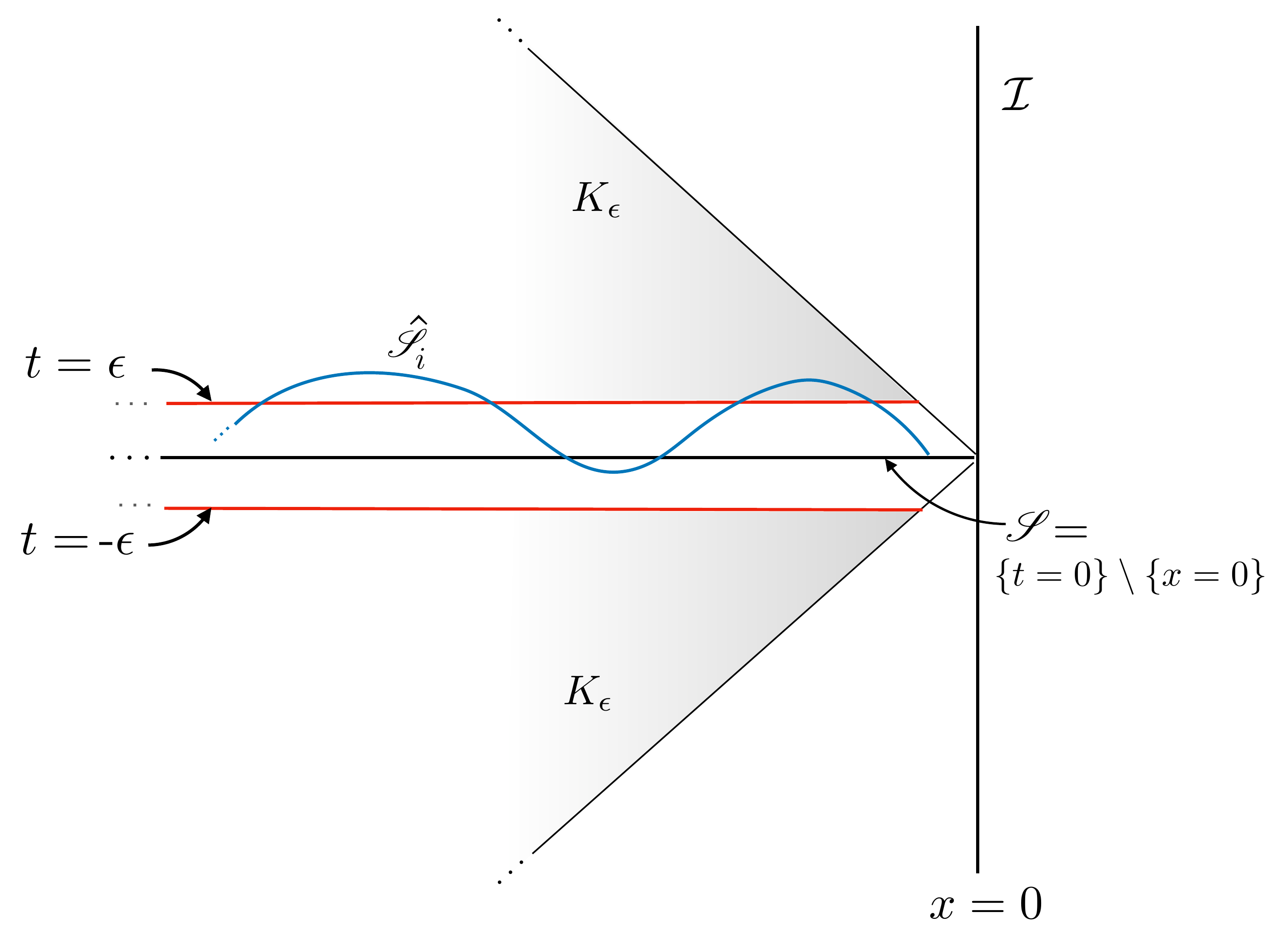}
\caption{The set $K_\epsilon$ for Bartnik's interior estimates.\label{F27VII22.1}}
\end{center}
\end{figure}

We return to the proof of Theorem~\ref{T11IX21.1}.  Let us denote by $u_\infty$  the limit  $\lim_{k\to\infty} u_{i_k}$.
It follows from the results of~\cite{bartnik:variational} that $u_\infty$ is smooth,
and its graph is spacelike.
 To see this, let $\epsilon>0$ and consider the compact set
$$
 K_\epsilon:=\overline{\mcD(\hyp,\mcM)\cap \{|t|\ge \epsilon\}}
  \,,
$$
see Figure~\ref{F27VII22.1}.
Since $\whyp_i$ is contained in
$   \overline{\mcD(\hyp_i,\mcM)} \subset \overline{\mcD(\hyp,\mcM)}
$,
the intersection of $\whyp_i$ with $\partial K_\epsilon$, if non-empty, is contained in $|t|=\epsilon$.
We can thus appeal to Theorem~3.1 of \cite{bartnik:variational} to obtain an $i$-independent,
 possibly $\epsilon$-dependent, estimate for the tilt function $\nu$ of $\,\whyp_i\cap\{|t|\ge 2\epsilon\}$.
 Elliptic estimates
show that the hypersurfaces $\whyp_{i_k}\cap\{|t| > 2\epsilon\}$ form a sequence which converges, as $k\to\infty$, in  $C^2_{\mathrm{loc}}$ to a smooth graph over
$\hyp\setminus\{ |u_\infty| > 2\epsilon\}  $
 with mean curvature $H$ equal to $H_1$. Since $\epsilon$ is arbitrary, we conclude that the graph of $u_\infty$ is smooth, except perhaps where it intersects $\{t=0\}$.

In order to handle that last case, let $p\in \mcM$ be such that $u_\infty(p)=0$. Let $\check t$ be a time function such that the zero-level set of $\check t$,  say $\check \hyp_0$, intersected with $\red{\tScri}$ lies to the timelike future of the zero-level set of $t$, and such that $\check \hyp_0$ lies to the timelike past of $p$ near $p$.   Theorem~3.1 of \cite{bartnik:variational} shows that the tilt function of $ \whyp_{i_k}\cap J^+(\check \hyp_0)$, defined  with respect to the time function $\check t$, is uniformly bounded near $p$, independently of $k$, and smoothness of $\whyp_\infty$ near $p$ follows.
As $p$ was arbitrary, we  conclude that $\whyp_\infty$ is smooth everywhere, and has  mean curvature  equal to $H_1$ everywhere.

Note that for every $j$ the graph of $u_\infty|_{\hyp_j}$ is spacelike, in particular it can neither intersect nor touch $\partial  \mcD(\hyp,\mcM)$. It then follows that the graph of $u_\infty$ is a Cauchy surface for  $\mcD(\hyp)$ by standard arguments (cf., e.g.,~\cite{BILY,Galloway:cauchy}).
\qed

\appendix
\section{Checking the hypotheses of Theorem~\ref{T19III22.1}}
 \label{App6IV22.1}

In this Appendix we check the hypotheses of Theorem~\ref{T19III22.1}, for metrics in the Fefferman-Graham form \eqref{9V21.6}-\eqref{9V21.7b}, in $(n+1)$-dimensions.

Near the conformal boundary $\partial \bmcM$, an orthonormal coframe $\theta^\mu$, both for $\fourg$ and for $\backriemg$, is given by
\begin{equation}\label{21III22.31}
  \theta^0 = x^{-1} {\hat \alpha}  dt
  \,,
  \quad
  \theta^A=  x^{-1}\hat \theta^A{}_{B}(dx^B + \hat \beta^B dt)
  \,,
  \quad
  \theta^n=  x^{-1}dx
  \,,
\end{equation}
where $\hat\theta^A{}_B dx^B$ is an orthonormal coframe for $h_{AB}dx^Adx^B$. Thus, to obtain the pointwise $\backriemg$-norm-squared of tensor we can find the components of this tensor in this frame, and calculate the sum of squares of these components.

Let $K$ be a compact  subset of the conformally completed manifold $\bmcM $.

In what follows we assume that $x^{2} \fourg$ extends   to a tensor field defined on  $\bmcM$ of $C^{2}$-up-to-boundary differentiability class.

\medskip

$\bullet $
$\Ric$:
The Ricci tensor of $\fourg$ asymptotes to
  $\frac{R}{n+1}\eta_{\alpha\beta}\theta^\alpha \theta^\beta$, where $R$ is the curvature scalar of $\fourg$ and $\eta_{\alpha\beta}$ is the Minkowski quadratic form. The bound
\begin{equation}\label{21III22.21a}
  |\Ric |_{\backriemg} \le C
  \,,
\end{equation}
on $K$ readily follows.

\medskip

$\bullet $
$ \alpha^{-1} \nabla \alpha $:
We have
\begin{eqnarray}
  \alpha^{-1} \nabla \alpha &   = & x^{-1} \nabla x
  +\hat \alpha^{-1} \nabla \hat \alpha
   \quad
   \Longrightarrow
   \quad
   |  \alpha^{-1} \nabla \alpha |_{\backriemg} = 1 + O(x)
   \,,
    \label{26III22.1}
\end{eqnarray}
which clearly implies a uniform bound on $ |  \alpha^{-1} \nabla \alpha |_{\backriemg}$ on every compact $K$.
\medskip

$\bullet $
$ \alpha^{-1} \nabla \nabla \alpha $:
Using the notation of \eqref{26III22.6} one finds
\begin{eqnarray}
  \alpha^{-1} \nabla \nabla \alpha &   = & 2 x^{-2} \nabla x \otimes \nabla x
  -  x^{-1} \nabla  \nabla x
  +\hat \alpha^{-1} \nabla  \nabla \hat \alpha
   \nonumber
\\
 &&
  - x^{-1}   \hat \alpha^{-1}
  ( \nabla x \otimes \nabla \hat \alpha + \nabla \hat \alpha
   \otimes \nabla x)
   \nonumber
\\ &   = &
    x^{-1}(  \nabla  \nabla x + \hat g)
  +\hat \alpha^{-1} \nabla  \nabla \hat \alpha
   \nonumber
\\
 &&
  - x^{-1}   \hat \alpha^{-1}
  ( \nabla x \otimes \nabla \hat \alpha + \nabla \hat \alpha
   \otimes \nabla x)
   \,,
    \label{26III22.1sea}
\end{eqnarray}
and a bound on $ |  \alpha^{-1} \nabla  \nabla \alpha |_{\backriemg}$ on compact sets follows from \eqref{26III22.4}-\eqref{26III22.5}, which give
\begin{eqnarray}
  |\alpha^{-1} \nabla \nabla \alpha|^2_{\backriemg} &   =
  &
   n + O(x)
   \,.
    \label{26III22.1seb}
\end{eqnarray}

\medskip

$\bullet $
$ {\mcL_T \fourg}$:
 Recall that
\begin{eqnarray}
  T =  - \alpha  \nabla t \equiv \alpha^{-1} ( \partial_t  - \beta^i \partial_i )\equiv x \hat\alpha^{-1} ( \partial_t  - \beta^i \partial_i )
   \,.
    \label{28IV21.4ag}
\end{eqnarray}
Since $\mcL_T (x^{-2} \chi) =  x^{-2}\mcL_T  \chi$ for any tensor field $\chi$, the estimate
\begin{equation}\label{21III22.21b}
  |{\mcL_T \fourg} |_{\backriemg} \le C
  \,,
\end{equation}
on $K$ is straightforward.

\medskip

$\bullet $
$ \nabla {\mcL_T \fourg}$:
From \eqref{28IV21.4ag} (recall that $\beta^x=0$) we have
\begin{eqnarray}
  \nabla {\mcL_T \fourg} &   = &  \nabla ( x^{-2} \mcL_T \overline{\fourg})
   \,.
    \label{28IV21.4agl}
\end{eqnarray}
Using \eqref{26III22.4}-\eqref{26III22.5} one finds
\begin{equation}\label{26III22.11}
  | \nabla {\mcL_T \fourg} |_{\backriemg} \le C
  \,,
\end{equation}
where $C$ might depend  upon $K$.
%
%

\medskip

$\bullet $
$ H_{\partial \Omega}$:
We consider the mean-curvature vector of the $(n-1)$-dimensional manifolds, say  $S_{t,x}$, obtained by intersecting the level sets of $x$ with the level sets of $t$. Let $\{\hat e_{A} = \hat e_A{}^B \partial_B\}_{A=2}^n$ be   ON bases for the metrics $h_{AB}dx^A dx^B$ induced by $\fourg$ on the $S_{t,x}$'s, then the vector fields $e_A = x \hat e_A$ are tangent to the $S_{t,x}$'s, and the collection $\{x\partial_x,T,e_A\}$ provides, at each point near the conformal boundary at infinity, an ON basis both for $\fourg $ and for $\backriemg$. By definition,
\begin{equation}\label{27III22.5}
  H_{S_{t,x}} : = - \sum_{A=2}^n (\nabla_{e_A}e_A)^\perp
  \equiv - x^2 \sum_{A=2}^n (\nabla_{\hat e_A}\hat e_A)^\perp
  \,,
\end{equation}
where $\perp$ denotes $\fourg$-orthogonal projection to $(T S_{t,x})^\perp$:
\begin{equation}\label{27III22.6}
  X^\perp = - \fourg(X,T)T +    \fourg(X,x\partial_x) x \partial_x
  \,.
\end{equation}
Equations~\eqref{26III22.4}-\eqref{26III22.5} give
\begin{equation}\label{27III22.6a}
    \fourg(X,T)= O(x)
    \,,
    \qquad
       \fourg(X,x\partial_x) = n-1 + O(x)
  \,,
\end{equation}
and boundedness of
$ | H_{S_{t,x}}|_{\backriemg} $, with a constant independent of $t$ and $x$ on any compact subset of $\R_t\times [0,x_0]_x$,
 follows.

\section{Christoffel symbols}
 \label{App26III22.1}

Consider a
metric $g$ of the form
\bel{26III22.6} g=x^{-2}\overline{g}=x^{-2}(dx^2+\hat{g}(x))\;,\qquad
\hat{g}(x)(\partial_x,\cdot)=0\;,
\ee
on
$[0,\epsilon]\times\partial \mcM $, where $\{\hat g(x)\}_{x\in[0,\epsilon]} $ is a collection of metrics on the level sets of $x$. Let $(x^A) = (x^{2},...,x^{{n}})$ be a
local coordinate system on $\partial \mcM $. We will work in the
coordinate system $(x^1 = r, x ^2, ... x^n)$. Then
\beal{26III22.4}
 &\Gamma^x_{xx}=-x^{-1}\,,\;\;
 \Gamma^A_{xx}=\Gamma^x_{Ax}=0\,,\;\;
  \Gamma^x_{AB}=x^{-1}\hat{g}_{AB}(x)-\frac{1}{2}\hat{g}'_{AB}(x)\,,
 &
\\
 & \Gamma^C_{xA}=-x^{-1}\delta^C_A+\frac{1}{2}\hat{g}^{CD}(x)\hat{g}'_{DA}(x)
 \,,
  \;\;
  \Gamma^C_{AB}=\hat{\Gamma}^{C}_{AB}(x)\,.
\eeal{26III22.5}
Here $f'$
denotes the derivative of a function $f$ with respect to $x$.

\section{Extending time functions}
 \label{app13VII22.1}

The aim of this appendix is to present a simple construction of a time function as needed in the proof of Theorem~\ref{T3IV22.2}.

Let $t$ denote the time coordinate in a Fefferman-Graham coordinate system in which \eqref{28III22.11} holds.

We let $\epsilon_0$ be as in the proof of Theorem~\ref{T3IV22.2}.
In the context of Theorem~\ref{T3IV22.2} the time function of interest will be defined
on the globally hyperbolic set $\mcD(\{t=0,0\le x\le \epsilon_0\},\ovmcM)$ with compact closure in $\ovmcM$.
For the purpose of Theorem~\ref{T11IX21.1} the time function constructed in this appendix is defined throughout $\omcM$.

By~\cite[Theorem 1.1 and Remark 4.14]{BernalSanchez2} there exists a time function, say $\hat t$, on $\mcD(\{t=0\}\cap \{0<x<\epsilon_0\},\mcM)$ so that its zero-level set, say  $\whhyp  $,
contains the hypersurfaces with boundary $\{t=0\,, \ x\in[\epsilon/4, \epsilon/2]\}$ and   $\hyp\cap\{x\in [\epsilon,2\epsilon]\}$, where we assume that $2\epsilon <\epsilon_0$.
Note that
$\hthyp  $ interpolates smoothly between the hypersurfaces $\{t=0\,,\, x\le \epsilon/2\}$
and $\hyp\cap \{x\ge \epsilon\}$. We will denote by $\wthyp$ the hypersurface so obtained.

Let $\fourg_1 $ be any smooth Lorentzian metric on $\bmcM$   which equals
\begin{equation}
 \label{16VII22.1}
 -dt^2+x^{2}g_{ij}dx^i dx^j
 \ \mbox{for $x\le \epsilon/4$}
\end{equation}
and for which $\wthyp  $ is spacelike. (For example, deform $\fourg$ slightly in the region  $x\le \epsilon/2$ to achieve \eqref{16VII22.1}, and leave it as it was elsewhere.)
Let  $\tau$ be the signed $\fourg_1 $-geodesic distance function from
$\wthyp  $, note that $\tau$ coincides with $t$ for $x\le \epsilon/4$. There exists $\delta >0$ so that $\tau$ is a smooth function with $\fourg$-timelike gradient for $p$ such that $|\tau (p)|< \delta$ and $x(p)\le \epsilon_0$.

 Let $\phi:\R\to\R$ be any smooth  non-decreasing function such that $\phi(z)=1$ for $z\ge \delta /2$,  with $\phi(z)=0$, and   $\phi(z)=-1$ for $z\le -\delta/2$, with $\phi'(z) \ne 0$ for $z\in (-\delta/2,\delta/2)$.
Then $\phi\circ \tau$ is a smooth semi-time function which vanishes on $\wthyp  $.

Let $t_+$ be any smooth semi-time function which vanishes for $\tau\le \delta/8$ and which is a time function  on $\tau\ge \delta/4$; such a function can be constructed e.g.\ using smoothed-out Geroch-type volumes as in~\cite{ChruscielGrantMinguzzi,Seifert,GerochDoD}.
Similarly let $t_-$ be any smooth semi-time function which vanishes for $\tau\ge -\delta/8$ and which is a
time function  on $\{\tau\le -\delta/4\}$. Then
$$
 \tilde  t := \phi \circ \tau + t_+ +t_-
$$
is a time function which vanishes on $\wthyp  $ and coincides with $t$ for $|t|\le \delta/4$ and $x\le \epsilon/4$.

\bigskip

{\noindent \sc Acknowledgements:}
The motivation for this paper was  a private question of Edward Witten.
Useful discussions with Rafe Mazzeo, Ettore Minguzzi and Edward Witten are acknowledged.
PTC is grateful to the University of Miami for hospitality.  Part of  his research was performed while visiting the Institute for Pure and Applied Mathematics (IPAM), which is supported by the National Science Foundation (Grant No. DMS- DMS-1925919).   The research of GJG was supported in part by the National Science Foundation (Grant No. DMS-171080) and by the Simons Foundation (Grant No. 850541).

\providecommand{\bysame}{\leavevmode\hbox to3em{\hrulefill}\thinspace}
\providecommand{\MR}{\relax\ifhmode\unskip\space\fi MR }
\providecommand{\MRhref}[2]{%
  \href{http://www.ams.org/mathscinet-getitem?mr=#1}{#2}
}
\providecommand{\href}[2]{#2}

\end{document}